\documentclass[prl,twocolumn,showpacs,preprintnumbers,superscriptaddress,longbibliography]{revtex4-2}

\usepackage{amsmath,amssymb}
\usepackage{graphicx}
\usepackage{mathrsfs}
\usepackage{dcolumn}
\usepackage{bm}
\usepackage{color}
\usepackage{bbm}
\usepackage[colorlinks=true,citecolor=blue,linkcolor=blue,urlcolor=blue]{hyperref}
\usepackage{svg}
\usepackage{soul}
\usepackage{rotating}

\begin{document}

\title{Entanglement R\'{e}nyi Negativity of Interacting Fermions \\ from Quantum Monte Carlo Simulations}
\author{Fo-Hong Wang}
\affiliation{Key Laboratory of Artificial Structures and Quantum Control (Ministry of Education), School of Physics and Astronomy \& Tsung-Dao Lee Institute, Shanghai Jiao Tong University, Shanghai 200240, China}
\author{Xiao Yan Xu}
\email{xiaoyanxu@sjtu.edu.cn}
\affiliation{Key Laboratory of Artificial Structures and Quantum Control (Ministry of Education), School of Physics and Astronomy \& Tsung-Dao Lee Institute, Shanghai Jiao Tong University, Shanghai 200240, China}
\affiliation{Hefei National Laboratory, Hefei 230088, China}
\date{\today}

\begin{abstract}
    Many-body entanglement unveils additional aspects of quantum matter and offers insights into strongly correlated physics. While ground-state entanglement has received much attention in the past decade, the study of mixed-state quantum entanglement using negativity in interacting fermionic systems remains largely unexplored. We demonstrate that the partially transposed density matrix of interacting fermions, similar to their reduced density matrix, can be expressed as a weighted sum of Gaussian states describing free fermions, enabling the calculation of rank-$n$ R\'{e}nyi negativity within the determinant quantum Monte Carlo framework. 
    We calculate the rank-two R\'{e}nyi negativity for the half-filled Hubbard model and the spinless $t$-$V$ model. Our calculation reveals that the area law coefficient of the R\'{e}nyi negativity for the spinless $t$-$V$ model has a logarithmic finite-size scaling at the finite-temperature transition point. 
    Our work contributes to the calculation of entanglement and sets the stage for future studies on quantum entanglement in various fermionic many-body mixed states.
\end{abstract}

\maketitle

\section{Introduction}

The characterization of emerging quantum many-body phenomena is multifaceted. Traditionally, physicists have relied on local measurements based on linear response to investigate matter. In recent decades, the utilization of quantum entanglement, a fundamental concept in quantum physics and a powerful tool in quantum information, has become pivotal in unveiling the additional aspects of quantum matter, including the identification of exotic phases and quantum criticality~\cite{Phys.Rev.Lett.2003Vidal,Rev.Mod.Phys.2008Amico,PhysicsReports2016Laflorencie}. 
A prominent example is the entanglement entropy (EE) used in bipartite ground-state entanglement studies~\cite{J.Phys.AMath.Theor.2009Calabrese}, 
where various corrections to the area law~\cite{Rev.Mod.Phys.2010Eisert} have been employed to classify quantum phases. These include
logarithmic corrections in the leading area-law term for 1D critical systems~\cite{J.Stat.Mech.Theor.Exp.2004Calabrese} and Fermi surfaces in generic dimensions~\cite{Phys.Rev.Lett.2006Gioev},
subleading logarithmic terms for corner contributions to 2D critical systems~\cite{Phys.Rev.Lett.2006Fradkin} and Goldstone modes in symmetry-breaking phases~\cite{Phys.Rev.B2011Kallin},
and the topological EE for non-local orders~\cite{Phys.Rev.Lett.2006Kitaev,Phys.Rev.Lett.2006Levin}.

However, EE is not a faithful mixed-state entanglement measurement due to its incompetence in distinguishing quantum entanglement from classical correlation. Thus, many entanglement measurements for mixed states have been proposed~\cite{QuantumInfo.Comput.2007Plbnio}, including the entanglement negativity~\cite{Phys.Rev.A1998Zyczkowski,J.Mod.Opt.1999Eisert,Phys.Rev.A2002Vidal,Phys.Rev.Lett.2005Plenio} (referred to as ``negativity'' henceforth for brevity), which was designed base on positive partial transpose criteria for the separability of density matrices~\cite{Phys.Rev.Lett.1996Peres,PhysicsLettersA1996Horodecki}. The evaluation of negativity hinges on the partial transpose of the given density matrix and can be carried out straightforwardly through basic matrix manipulations without invoking any optimization. Hence, negativity has been employed to examine entanglement in finite-temperature Gibbs states or tripartite ground states in various systems, spanning from one-dimensional conformal field theory~\cite{Phys.Rev.Lett.2012Calabrese,J.Stat.Mech.2013Calabrese,J.Phys.AMath.Theor.2014Calabrese}, bosonic systems~\cite{Phys.Rev.A2002Audenaert,Phys.Rev.B2014Chunga,NewJ.Phys.2014Eisler,Phys.Rev.B2016Eisler}, spin systems~\cite{Phys.Rev.Lett.2012Bayat,J.Stat.Mech.2013Calabresea,J.Stat.Mech.2013Alba,Phys.Rev.E2016Sherman,Phys.Rev.Lett.2020Wu}, to topologically ordered phases~\cite{Phys.Rev.A2013Castelnovo,Phys.Rev.A2013Lee,Phys.Rev.B2019Lu,Phys.Rev.Lett.2020Lu,Phys.Rev.Research2020Lu}.

In the case of fermionic systems, the definition of partial transpose needs to be adjusted to accommodate the anticommuting statistical property. There exist two different proposals for fermionic partial transpose (FPT) and corresponding fermionic negativity, as discussed in Refs.~\cite{NewJ.Phys.2015Eisler,Phys.Rev.B2018Eisert} and Refs.~\cite{Phys.Rev.B2017Shapourian,Phys.Rev.B2018Shiozaki,Phys.Rev.A2019Shapourian} respectively. Despite being a computable entanglement measurement, fermionic negativity is only analytically tractable in free systems, especially at finite temperatures, and there have been studies based on both the former definition~\cite{Phys.Rev.B2016Chang,Phys.Rev.B2016Eisler,J.Stat.Mech.2016Coser,J.Stat.Mech.2016Herzog} and the latter definition~\cite{Phys.Rev.B2017Shapourian,J.Stat.Mech.2019Shapourian,SciPostPhys.2023Alba}. Therefore, it is desirable to design a quantum Monte Carlo (QMC) algorithm for large-scale simulation of interacting fermionic systems in an unbiased manner, which is the main goal of this work. Throughout this paper, we adopt the definition in Refs.~\cite{Phys.Rev.B2017Shapourian,Phys.Rev.B2018Shiozaki} under which the partial transpose of a Gaussian state remains a Gaussian state. Additionally, instead of utilizing the originally proposed negativity which involves trace norm of partially transposed density matrices (PTDMs)~\cite{Phys.Rev.A2002Vidal}, we consider R\'{e}nyi negativity (RN) which involves moments of PTDMs, as done in several previous studies on other systems~\cite{Phys.Rev.Lett.2012Calabrese,J.Stat.Mech.2013Calabrese,J.Stat.Mech.2013Alba,Phys.Rev.B2014Chunga,Phys.Rev.B2016Eisler,Phys.Rev.Lett.2020Wu}. 

In fact, our main result is more broadly applicable. We show that generic PTDMs can be written as a weighted sum of Gaussian states, representing free fermions coupled with auxiliary fields, similar to Grover's pioneering work on reduced density matrices for EE~\cite{Phys.Rev.Lett.2013Grover}. Our finding facilitates the calculation of RN in a tractable manner, thus establishing it as a powerful tool for characterizing entanglement in mixed states of interacting fermions.
We demonstrate this relation using determinant quantum Monte Carlo (DQMC) simulations~\cite{Phys.Rev.D1981Blankenbecler,Phys.Rev.B1981Scalapino,Assaad2008WorldlineDeterminantalQuantum} on two paradigmatic models in the realm of strongly correlated electrons, namely, the Hubbard model and the spinless $t$-$V$ model. These two models on bipartite lattices at half-filling are sign-problem-free and both ground-state and finite-temperature properties can be feasibly simulated within the DQMC framework. The relation between negativity and finite temperature transition in fermionic systems is unveiled. 

\section{Results}
\subsection{Partially transposed density matrix integrated to DQMC framework}

Various definitions of negativity in the literature share a common and central dependency, namely, the partial transpose of the density matrix. In this work, we adopt the partial time-reversal transformation proposed by Shapourian \textit{et al.}~\cite{Phys.Rev.B2017Shapourian,Phys.Rev.B2018Shiozaki} as the FPT.

\begin{figure}
    \centering
    \includegraphics[width=\columnwidth]{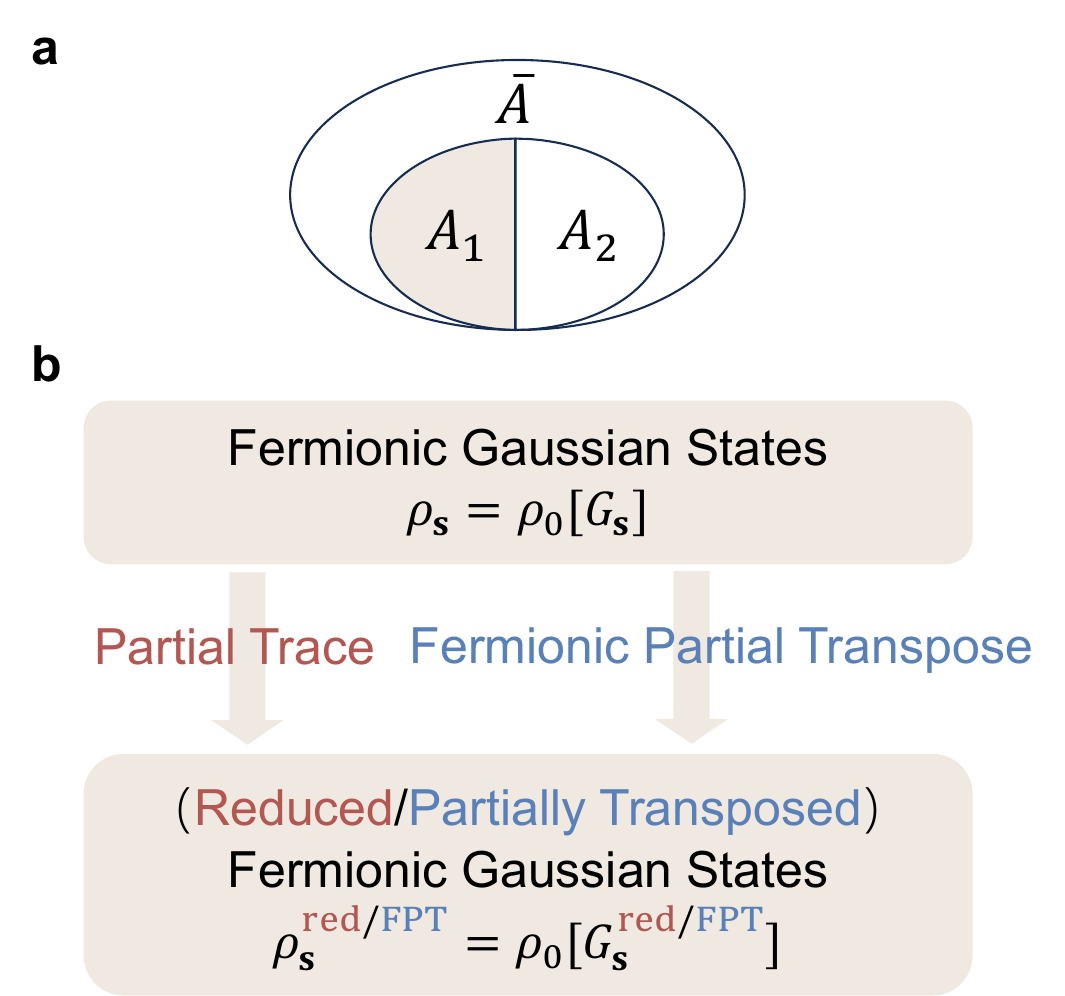}
    \caption{
    A schematic illustration of the core concepts of this work.
    (a) Illustration of the general tripartite geometry. First, we trace out the environment $\bar{A}$ to obtain the reduced density matrix $\rho_A$, and then evaluate the entanglement between subsystems by either tracing out or partially transposing $A_2$. 
    (b) The fermionic partial transpose of Gaussian states, $\rho_0[G]\sim e^{\mathbf{c}^\dagger\ln(G^{-1}-I)\mathbf{c}}$, remains Gaussian. These states represent free fermions, including the auxiliary-field-dependent density matrix $\rho_\mathbf{s}$ in the DQMC framework. 
    The reduced and partially transposed density matrices thus share a unified expression in terms of the corresponding Green's functions. 
    Specifically, the reduced Green's function is $G^{\rm red}_{\mathbf{s},A_1}=\langle c_jc_k^\dagger\rangle_\mathbf{s}$ for $j,k\in A_1$, while the partially transposed Green's function $G_{\mathbf{s},A_1}^{\rm FPT}=G_\mathbf{s}^{T_2^f}$ is given by Eq. \eqref{equ:FPT_Green}. 
    }
    \label{fig:descreption}
\end{figure}

We begin with the general partitioning of a fermionic lattice model. It is defined using annihilation (creation) operators $c_{j\sigma}^{(\dagger)}$, which satisfy the anticommutation relations $\{c_{j\sigma}, c^\dagger_{k\sigma^\prime}\}=\delta_{jk}\delta_{\sigma\sigma^\prime}$, where $j,k=1,\dots,N$ are the labels of the sites and $\sigma,\sigma^\prime$ are the indices for internal degrees of freedom such as spin. 
In the following discussion, we may use a column vector $\mathbf{c}=(c_{1,\uparrow},\dots,c_{N,\uparrow},c_{1,\downarrow},\dots,c_{N,\downarrow})^T$ to compactly encapsulate all the fermionic operators. 
This lattice system, denoted as $A$, generally exists within a larger space, as illustrated in Fig.~\ref{fig:descreption}A. After tracing out the environment $\bar{A}$, system $A$ typically exists in a mixed state $\rho$. For example, if system $A$ is in contact with a much larger thermal bath at temperature $T$, then we obtain a finite-temperature Gibbs state $\rho=e^{-\beta H}/\mathrm{Tr}e^{-\beta H}$ with $\beta=1/T$ the inverse temperature and $H$ the Hamiltonian of the system $A$. 
Next, we further divide system $A$ into two parties belonging to two complementary spatial regions respectively, i.e., $A=A_1\cup A_2$. Then the density matrix acting on Hilbert space $\mathcal{H}_1\otimes \mathcal{H}_2$ can be expanded as $\rho=\sum_{A_1,A_2,A_1^\prime,A_2^\prime}\rho_{A_1,A_2;A_1^\prime,A_2^\prime}|A_1\rangle|A_2\rangle\langle A_1^\prime| \langle A_2^\prime|$. 

The FPT of density matrix $\rho$ with respect to subsystem $A_2$, denoted as $\rho^{T_2^f}$, exhibits a highly succinct mathematical expression in the Majorana basis~\cite{Phys.Rev.B2017Shapourian,Phys.Rev.B2018Shiozaki}. 
Under Majorana basis, an arbitrary density operator can be expressed as a constrained superposition of products of Majorana operators, which are defined as $\gamma_{2j-1,\sigma}=c_{j,\sigma}+c^\dagger_{j,\sigma}$ and $\gamma_{2j,\sigma}=-\mathrm{i}(c_{j,\sigma}-c_{j,\sigma}^\dagger)$. 
It is found that $\rho^{T^f_2}$ can be obtained by applying the following transformation to the Majorana operators associated with subsystem $A_2$:
\begin{equation}\label{equ:FPT_majorana}
    \mathcal{R}_2^f(\gamma_{j,\sigma})=\text{i}\gamma_{j,\sigma},\quad j\in A_2.
\end{equation}
Remarkably, under this definition, the partial transpose of a Gaussian state, denoted as 
$\rho_0 \sim e^{\frac{1}{4}\boldsymbol\gamma^{T}W\boldsymbol\gamma}$ with $\boldsymbol\gamma=(\gamma_{1,\uparrow},\dots,\gamma_{2N,\uparrow},\gamma_{1,\downarrow},\dots,\gamma_{2N,\downarrow})^T$,
retains its Gaussian nature. 
The question then pertains to determining the explicit form of $\rho_0^{T_2^f}$ or $W^{T_2^f}$.
To this end, it is important to emphasize that a Gaussian state $\rho_0$ can be alternatively characterized by the Green's function $\Gamma_{kl}=\langle[\gamma_{k},\gamma_{l}]\rangle/2$, which is averaged with respect to $\rho_0$ itself and also called covariance matrix. 
This matrix is connected to the $W$ matrix through the relation $\tanh(-W/2)=\Gamma$, or inversely, $W=\ln\left[(I+\Gamma)^{-1}(I-\Gamma)\right]$~\cite{J.Stat.Mech.2010Fagotti} (see also the Supplementary Note 3 for a proof). 
By employing the definition of $\Gamma$ and the partial transpose in the Majorana basis (refer to Eq.~\eqref{equ:FPT_majorana}), the partial transpose of the covariance matrix can be formulated as
\begin{equation}\label{equ:FPT_Gamma}
    \Gamma^{T_{2}^{f}}=\left(\begin{array}{cc}
        \Gamma^{11} & \text{i}\Gamma^{12}\\
        \text{i}\Gamma^{21} & -\Gamma^{22}
        \end{array}\right),
\end{equation}
where $\Gamma^{bb^\prime}$ ($b,b^\prime =1,2$) denotes the block comprising the matrix elements with rows pertaining to subsystem $A_b$ and columns pertaining to subsystem $A_{b^\prime}$.
The Gaussian state described by $\Gamma^{T_{2}^{f}}$ precisely yields $\rho_0^{T_2^f}$ through the relation $\tanh(-{W^{T_2^f}}/{2})=\Gamma^{T_{2}^{f}}$, i.e., $(\rho_0[\Gamma])^{T_2^f}=\rho_0[\Gamma^{T^f_2}]$ with $\rho_0[\Gamma]\sim e^{\boldsymbol{\gamma}^T \ln\left[(I+\Gamma)^{-1}(I-\Gamma)\right]\boldsymbol{\gamma}}$.
This elegant fact is proved using Wick's theorem for Majorana monomials~\cite{NewJ.Phys.2015Eisler} (see the Supplementary Note 3 for details). 

The above discussion in the Majorana basis can be seamlessly transitioned to the complex fermion basis.
In complex fermion basis, the Green's function is defined as $G_{jk}=\langle c_{j}c_{k}^{\dagger}\rangle$, 
where we have abbreviated the spin indices. 
Its partially transposed form exhibits also a simple structure
\begin{equation}\label{equ:FPT_Green}
    G^{T_{2}^{f}}=\left(\begin{array}{cc}
        G^{11} & \text{i}G^{12}\\
        \text{i}G^{21} & I-G^{22}
        \end{array}\right),
\end{equation}
where the superscripts of the blocks $G^{bb^\prime}$ indicate the subsystems, akin to the notation of $\Gamma^{bb^\prime}$ established earlier.
Similar to the Majorana basis, the above Green's function delineates another Gaussian state which is exactly the partial transpose of the original Gaussian state, i.e., $(\rho_0[G])^{T_2^f}=\rho_0[G^{T^f_2}]$ with $\rho_0[G]\sim e^{\mathbf{c}^\dagger\ln(G^{-1}-I)\mathbf{c}}$~\cite{Phys.Rev.B2004Cheong}.

It is now pertinent to redirect our attention towards the partial transpose for interacting fermionic systems, whose density matrices are not Gaussian states. 
Nonetheless, within the framework of DQMC, the original Hamiltonian $H$ is transformed by replacing interaction terms with fermion bilinears coupled to spacetime-dependent auxiliary fields $\mathbf{s}$~\cite{Phys.Rev.D1981Blankenbecler,Phys.Rev.B1981Scalapino,Assaad2008WorldlineDeterminantalQuantum}. 
Specifically, we consider the Gibbs state $\rho=e^{-\beta H}/Z$, where $H$ consists of a free-fermion term $H_0$ and a two-particle interaction term $H_I$. 
We employ Trotter decomposition to factorize the exponential operator $e^{-\beta H}$ as $(e^{-\Delta_{\tau}H})^{L_{\tau}}=\prod_{l=1}^{L_{\tau}}e^{-\Delta_{\tau}H_{I}}e^{-\Delta_{\tau}H_{0}}+O(\Delta_{\tau}^{2})$ with $L_\tau=\beta/\Delta_\tau$ being the number of imaginary-time slices. 
We then apply a Hubbard-Stratonovich (HS) transformation to decouple the interaction terms across different time slices. This procedure yields
\begin{equation}
    \rho=\frac{1}{Z}\sum_{\mathbf{s}}\eta\left[\mathbf{s}\right]\prod_{l=1}^{L_{\tau}}\left(e^{\mathbf{c}^{\dagger}V\left[\mathbf{s}(l)\right]\mathbf{c}}e^{-\Delta_{\tau}H_{0}}\right)\equiv\sum_{\mathbf{s}}P_{\mathbf{s}}\rho_{\mathbf{s}},
\end{equation}
where each $\mathbf{s}$-configuration is distributed over both the imaginary-time and spatial directions, contributing a scalar factor $\eta\left[\mathbf{s}\right]$ and a product of Gaussian operators.
Here, both $V[\mathbf{s}]$ and $\eta[\mathbf{s}]$ are derived from $H_I$, and their forms depend on the specific interactions and HS decoupling channels. For detailed expressions related to the two models examined in this study, please refer to the Supplementary Note 2. 
Since the product of Gaussian states remains a Gaussian state up to a normalization factor~\cite{J.Stat.Mech.2010Fagotti}, the interacting fermionic density matrix $\rho\sim e^{-\beta H}$ can ultimately be written as a weighted sum of Gaussian operators $\rho_\mathbf{s}$, with $P_\mathbf{s}$ denoting the configuration weight~\cite{Phys.Rev.Lett.2013Grover}. 
By leveraging the linearity of the partial transpose, we can first individually compute the FPT for each Gaussian state $\rho_\mathbf{s}$. 
We then sum these results, weighted by their respective probabilities $P_\mathbf{s}$, to obtain the FPT of the entire density matrix $\rho$:
\begin{equation}\label{equ:FPT_rhocomplex}
    \rho^{T_{2}^{f}}=\sum_{\bf{s}}P_{\bf{s}}\rho_{\bf{s}}^{T_{2}^{f}},
\end{equation}
where
\begin{equation}\label{equ:FPT_rhos}
    \rho_{{\bf {s}}}^{T_{2}^{f}}=\det\left[G_{{\bf {s}}}^{T_{2}^{f}}\right]\exp\left\{{\mathbf{c}^{\dagger}\ln\left[\left(G_{\mathbf{s}}^{T_{2}^{f}}\right)^{-1}-I\right]\mathbf{c}}\right\}.
\end{equation}
The aforementioned equations \eqref{equ:FPT_rhocomplex} and \eqref{equ:FPT_rhos}, along with Eq.~\eqref{equ:FPT_Green}, are the main results of this work and can be employed to investigate negativity and negativity spectrum within the conventional DQMC framework, fully analogous to the analysis of EE and entanglement spectrum (see Fig.~\ref{fig:descreption}B), respectively. 

\subsection{Quantum-classical crossover in Hubbard chain}

\begin{figure*}
    \centering
    \includegraphics[width=\textwidth]{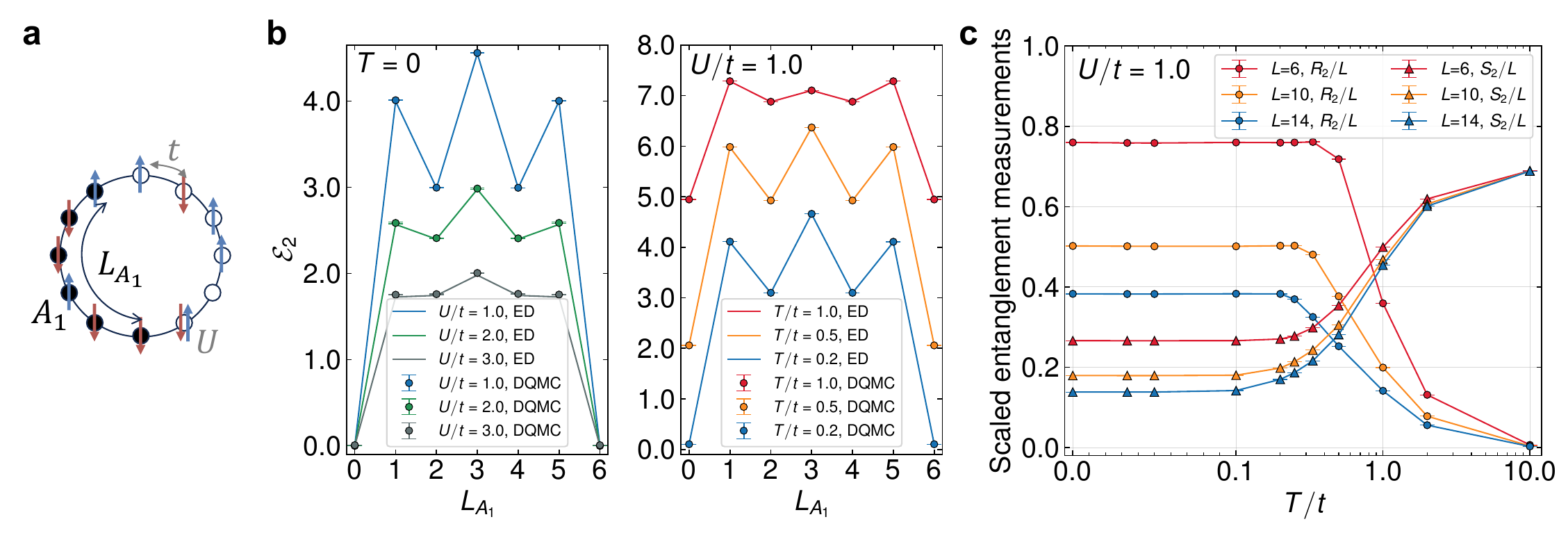}
    \caption{
    Rank-two RN(R) in the half-filled Hubbard chain.
    (a) A schematic illustration of the half-filled Hubbard chain with periodic boundary conditions and its bipartition geometry. 
    (b) The variation of the rank-two RN $\mathcal{E}_2$ for a six-site Hubbard chain is depicted as a function of the subsystem length $L_{A_1}$. The solid lines represent the ED results, which agree with the DQMC results at both zero temperature (left panel) and finite temperatures (right panel).
    (c) Quantum-classical crossover. The scaled RNR $R_2/L$ and EE $S_2/L$ of the half-filled Hubbard chain under a half-chain bipartition (i.e., we take $L_{A_1}=L/2$) vary as functions of temperature.
    As the temperature rises, the scaled EE for different lengths increases and converges, indicating a dominance of volume law at high temperatures. Meanwhile, the RNR begins to vanish once the temperature reaches a critical value associated with the finite-size gap $1/L$~\cite{J.Stat.Mech.2019Shapourian}. 
    The error bars in (b) and (c) represent the standard errors from Monte Carlo sampling.
    }
    \label{fig:HubbardChain}
\end{figure*}

We first consider the half-filled Hubbard chain with periodic boundary conditions, illustrated in Fig.~\ref{fig:HubbardChain}A and described by the Hamiltonian
\begin{equation}\label{equ:Hubbard}
    H=-t\sum_{\langle ij\rangle \sigma} (c^\dagger_{i\sigma} c_{j\sigma}+\mathrm{H.c.})+\frac{U}{2}\sum_{i}(n_{i}-1)^2, 
\end{equation}
which is a sign-problem-free model~\cite{Assaad2008WorldlineDeterminantalQuantum}. We will benchmark DQMC results from two perspectives: (i) a numerical comparison with results obtained from exact diagonalization (ED) where we employ the definition of FPT in the Fock space (see Supplementary Eqs. (2) and (3)), and (ii) providing a physical explanation for why the negativity is a more competent mixed-state entanglement measurement compared to EE in the context of a quantum-classical crossover~\cite{Phys.Rev.B2017Shapourian,J.Stat.Mech.2019Shapourian}. 

We define the rank-$n$ RN as
\begin{equation}\label{equ:nthRenyi_nega}
    \mathcal{E}_{n}=-\frac{1}{n-1}\ln\text{Tr}\left[\left(\rho^{T_{2}^f}\right)^{n}\right],
\end{equation}
where the $n$-th moment of the PTDM, denoted as $\text{Tr}[(\rho^{T_{2}^f})^{n}]$
, is also referred to as the replica approach of negativity in previous studies~\cite{Phys.Rev.Lett.2012Calabrese,J.Stat.Mech.2013Calabrese}.
The quantity $\mathcal{E}_n$ is formally a direct analog to rank-$n$ R\'{e}nyi EE $S_{n}(A_1)=-(\ln \mathrm{Tr}\rho^n_{A_1})/(n-1)$, where $\rho_{A_1}=\mathrm{Tr}_{A_2}\rho$ represents the reduced density operator obtained after tracing out subsystem $A_2$. Utilizing Eq.~\eqref{equ:FPT_rhocomplex}, we can derive the DQMC expression for measuring, for instance, the rank-two RN
\begin{equation}\label{equ:2thRenyi_nega_DQMC}
    \begin{aligned}
        \mathcal{E}_{2}=&-\ln\Big\{\sum_{{\bf s}_{1}{\bf s}_{2}}P_{{\bf s}_{1}}P_{{\bf s}_{2}}
        \det\left[G_{\bf{s}_{1}}^{T_{2}^{f}}G_{\bf{s}_{2}}^{T_{2}^{f}}+\left(I-G_{\bf{s}_{1}}^{T_{2}^{f}}\right)\left(I-G_{\mathbf{s}_{2}}^{T_{2}^{f}}\right)\right]\Big\}.
    \end{aligned}
\end{equation}
The distinction between the FPT and the conventional one is presented herein. For bosonic systems, considering that $\mathrm{Tr}[(\rho^{T_2})^2]=\mathrm{Tr}[\rho^2]$, the rank-two RN becomes trivial, thereby rendering the minimal meaningful rank as three~\cite{Phys.Rev.Lett.2012Calabrese,J.Stat.Mech.2013Calabrese,Phys.Rev.Lett.2020Wu}. 
However, we show in the occupation number representation that all fermionic PTDMs satisfy ${\rm Tr} [ (\rho^{T_{2}^{f}} )^{2} ]={\rm Tr}[ (\rho\hat{X}_{2} (\pi ))^2]$ with $\hat{X}_{2}\left(\theta\right)=e^{{\rm i}\theta\sum_{j\in A_{2}}{n}_{j}}$ being the disorder operator (see the Supplementary Note 1 for details). Consequently, $\mathcal{E}_2$ can reveal the entanglement information of the system.
As shown in Fig.~\ref{fig:HubbardChain}B, the results calculated by DQMC and ED show strong agreement in both the zero-temperature and the finite-temperature regimes. In the former regime, the pattern of the rank-two RN exhibits analogous variations to those of the rank-two R\'{e}nyi EE~\cite{Phys.Rev.Lett.2013Grover,J.Stat.Mech.2014Broecker}, in response to alterations in the length of subsystem $A_1$, denoted as $L_{A_1}$. However, at finite temperatures, the negativity maintains a symmetric pattern, which is different from the behavior of EE~\cite{Phys.Rev.Lett.2014Wang,J.Stat.Mech.2014Broecker}. As the temperature rises, the magnitude of the negativity increases, resulting in an overall non-zero shift corresponding to a non-zero thermodynamic entropy of $-\ln(\mathrm{Tr}\rho^2)$. 

Based on the above observation at finite temperatures, we also examine the ratio between $\text{Tr}[(\rho^{T_{2}^f})^{n}]$ and $\text{Tr}[\rho^{n}]$ dubbed the R\'{e}nyi negativity ratio (RNR)~\cite{J.Stat.Mech.2013Alba,Phys.Rev.B2014Chunga,Phys.Rev.Lett.2020Wu}
\begin{equation}\label{equ:nthRenyi_ratio}
    R_{n}=-\frac{1}{n-1}\ln\left\{\frac{\text{Tr}\left[\left(\rho^{T_2^f}\right)^{n}\right]}{\text{Tr}[\rho^{n}]}\right\}=\mathcal{E}_n-S^{\rm th}_n,
\end{equation}
where $S^{\rm th}_n=-(\ln \mathrm{Tr}\rho^n)/(n-1)$ denotes the thermodynamic R\'{e}nyi entropy, which equals $\mathcal{E}_n$ for either $A_1=A$ or $A_2=A$.
A faithful description of mixed-state entanglement necessitates the exclusion of the thermodynamic R\'{e}nyi entropy $S_n^{\rm th}$. 
In Fig.~\ref{fig:HubbardChain}C, we display the variations of the RNR and EE with temperature for three distinct lengths, namely $L=6,10,14$. Here, the subsystem $A_1$ is chosen to be half of the chain, yielding an equal bipartition. As the temperature rises, the EE increases while the RNR asymptotically diminishes to zero for all lengths. 
This serves as a compelling physical demonstration of the quantity $R_n$. In a generic mixed state, both quantum and classical correlations are present, and effective measurement of mixed-state entanglement should exclusively isolate the quantum correlations~\cite{Phys.Rev.A2002Vidal}. In the specific context of finite-temperature Gibbs states, the classical correlation is simply the thermal fluctuations delineated by the thermodynamic entropy $S_n^{\rm th}$.
Furthermore, at sufficiently low temperatures, the RNR remains constant and establishes a plateau, the length of which is associated with the finite-size gap $1/L$~\cite{J.Stat.Mech.2019Shapourian}. As depicted in Fig.~\ref{fig:HubbardChain}C, it is evident that with an increase in chain length, the plateau becomes narrower. In summary, the monotonic decay of the RNR with rising temperature signifies a crossover from a quantum entangled state to a classical mixed state. 

\subsection{Finite temperature transition in $t$-$V$ model.}

\begin{figure*}
    \centering
    \includegraphics[width=\textwidth]{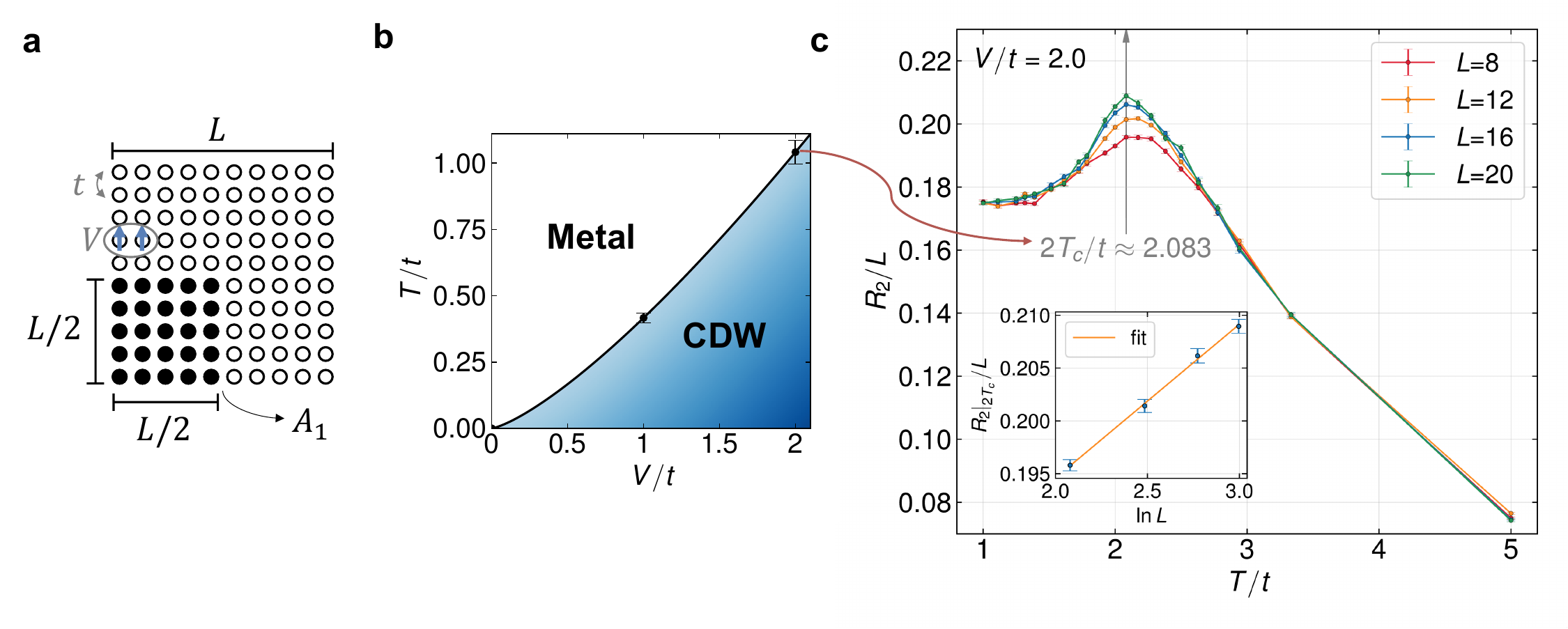}
    \caption{
        Rank-two RNR in the $t$-$V$ model across a finite-temperature transition point.
        (a) A schematic illustration of the spinless (or spin-polarized) $t$-$V$ model on a square lattice and the chosen one-quarter bipartition geometry. 
        (b) The phase diagram of the $t$-$V$ model, where the phase boundary points at $V=1$ and $V=2$ are determined through mixed-state entanglement studies. One can find the data of $V=1$ case in the Supplementary Fig.~\ref{fig:HubbardChain}. The error bars for these two points are estimated based on the difference between the neighboring temperature points that were calculated. 
        (c) The finite-temperature transition in the spinless $t$-$V$ model is detected by the area-law coefficient of the RNR as a function of temperature. A gray vertical arrow indicates the position of the shared peak, with half of its magnitude aligning with the transition point determined in previous studies~\cite{Phys.Rev.B1985Gubernatis,Phys.Rev.Lett.2014Wang}. The inset shows the linear scaling of the area-law coefficient at the critical point with $\ln L$. 
        The error bars in (c) represent the standard errors from Monte Carlo sampling.
        }
    \label{fig:Negaraio-tV}
\end{figure*}

To demonstrate the efficacy of the RNR in detecting finite-temperature phase transition, we further consider the half-filled spinless $t$-$V$ model on a square lattice with periodic boundary conditions~\cite{Phys.Rev.B1984Scalapino,Phys.Rev.B1985Gubernatis,Phys.Rev.Lett.2014Wang}, 
\begin{equation}
    H=-t\sum_{\langle i,j\rangle}(c_{i}^{\dagger}c_{j}+c_{j}^{\dagger}c_{i})+V\sum_{\langle i,j\rangle}\left(n_{i}-\frac{1}{2}\right)\left(n_{j}-\frac{1}{2}\right), 
\end{equation}
where both the hopping and the interaction involve only nearest neighbors (see Fig.~\ref{fig:Negaraio-tV}A for illustration).
In the presence of a finite coupling parameter $V$, this model exhibits a charge density wave (CDW) ground state and undergoes a phase transition from the CDW phase to a metallic phase at finite temperature (see Fig.~\ref{fig:Negaraio-tV}B), with critical behavior falling within the 2D Ising universality class~\cite{Phys.Rev.B1985Gubernatis,Phys.Rev.B2016Hesselmann}. In the following, we focus on a specific coupling strength, $V/t=2$, where the critical temperature was estimated to be $T_c/t\approx 1.0$~\cite{Phys.Rev.B1985Gubernatis}.

This model is also a sign-problem-free model~\cite{Phys.Rev.B2014Huffman,Phys.Rev.B2015Li,Phys.Rev.Lett.2015Wang,Phys.Rev.Lett.2016Wei}.
However, for models with larger dimensions or stronger interaction strengths, the direct sampling of RN using Eq.~\eqref{equ:2thRenyi_nega_DQMC} becomes inaccurate, as a result of the occurrence of spikes~\cite{Phys.Rev.E2016Shi} or the non-Gaussian distribution of Grover determinants $\det g_x=\det[G_{\bf{s}_{1}}^{T_{2}^{f}}G_{\bf{s}_{2}}^{T_{2}^{f}}+(I-G_{\bf{s}_{1}}^{T_{2}^{f}})(I-G_{\mathbf{s}_{2}}^{T_{2}^{f}})]$~\cite{ArXiv2023Liaoa}.
We implement an incremental algorithm for the RN, analogous to the controllable incremental algorithm for EE~\cite{Phys.Rev.Lett.2024DEmidio,Phys.Rev.B2023Pan,ArXiv2023Liaoa}, the spirit of which is to measure $(\det g_x)^{1/N_{\rm inc}}$ instead of $(\det g_x)$ to circumvent the sampling of an exponentially small quantity with exponentially large variance (see Methods). It is important to note that there is a sign ambiguity in the $N_{\rm inc}$-th root. In the Supplementary Note 5, we prove that the Grover determinant $\det g_x$ is always real and non-negative for two classes of sign-free models, represented by the Hubbard model and the spinless $t$-$V$ model, respectively. 

As illustrated in Fig.~\ref{fig:Negaraio-tV}A, we designate the lower left corner with dimensions $(L/2)\times(L/2)$ as subsystem $A_1$, resulting in an area-law coefficient of the RNR of $R_2/L$. 
The main plot of Fig.~\ref{fig:Negaraio-tV}C presents $R_2/L$ as a function of temperature for various system sizes, demonstrating a notably distinct finite-size characteristic compared to the intersection of mutual information~\cite{Phys.Rev.Lett.2011Singh,Phys.Rev.Lett.2014Wang}. 
Remarkably, unlike the Hubbard model in Fig.~\ref{fig:HubbardChain}C or the previous study on the $2+1\mathrm{D}$ transverse field Ising model~\cite{Phys.Rev.Lett.2020Wu}, the RNR does not exhibit a monotonic decrease with rising temperature. 
Instead, for varying lattice sizes, a shared local maximum appears at approximately twice the transition temperature, $2T_c/t\approx 2.1$. 
The inclusion of the prefactor $2$ aligns with the rank of the RNR under consideration, consistent with earlier discussion on the critical behavior within replica approach~\cite{Phys.Rev.B2019Lu,Phys.Rev.Research2020Lu,Phys.Rev.Lett.2020Wu}. 
In general, the singularity of $R_n$ is anticipated to occur at $T=nT_c$ with $T_c$ being the physical transition temperature. 
This expectation arises because $R_n$, as indicated by the denominator in its definition in Eq.~\eqref{equ:nthRenyi_ratio}, i.e., $\mathrm{Tr}[e^{-n\beta H}]$, effectively corresponds to a Gibbs state with an effective inverse temperature of $n\beta$.
Therefore, we demonstrate that it is possible to quantitatively extract the finite-temperature transition points and the phase diagram of the $t$-$V$ model (see Fig.~\ref{fig:Negaraio-tV}B) from mixed-state entanglement studies.
Finally, we briefly highlight the finite-size scaling of the rank-two RNR. As shown in the inset of Fig.~\ref{fig:Negaraio-tV}C, the peaks of $R_2/L$ exhibit a logarithmic divergence with system size $L$, while the area law is well-preserved in regions far from the critical point. 
This beyond-area-law scaling around the finite temperature critical point is also observed for other values of $V$ (namely, $V=0$ and $V=1$), a different bipartite geometry, and a different lattice. Refer to the Supplementary Fig.~\ref{fig:HubbardChain} for additional complementary plots.

\section{Discussion}

We showed that the PTDM for interacting fermions, akin to the reduced density matrix, can be expanded as a weighted sum of Gaussian states representing free fermions, thereby paving the way for the study of mixed-state entanglement in strongly correlated fermionic systems. 
This main result was employed to implement an algorithm to compute the rank-$n$ RN for interacting fermionic systems within the DQMC framework. We studied the rank-two RN for the half-filled Hubbard chain and the spinless $t$-$V$ model on a square lattice. 
Remarkably, we found that the area law coefficient of the RNR exhibits a logarithmic singular peak at about twice the finite-temperature transition point for all lattice sizes under consideration.

We now discuss the possible physical interpretations of the logarithmic divergence of the rank-two RNR at $2T_c$. 
Based on symmetry considerations, it was argued that the entanglement negativity inherits the singularity of the specific heat at a finite temperature transition~\cite{Phys.Rev.Research2020Lu,Phys.Rev.Lett.2020Wu}, and for the 2D Ising transition, the specific heat has a logarithmic divergence in lattice linear size. 
However, in our fermionic scenario, the quantity showing this divergence is $R_2/L$ rather than its temperature derivative. Thus it can not be directly connected to specific heat and the underlying cause of the logarithmic divergence of $R_2/L$ at $2T_c$ remains an open issue. We note that Refs.~\cite{Phys.Rev.Research2020Lu,Phys.Rev.Lett.2020Wu} concerning bosonic models used the conventional partial transpose, which may partly account for the inconsistency. 

There are several potential future research directions to consider.
The first direction is to investigate the finite-temperature entanglement of various interacting fermionic models, especially those with transition points that belong to different universality classes, such as the 3D Hubbard model which owns a transition belonging to $\mathrm{O}(3)$ universality class~\cite{Phys.Rev.Lett.2011Rohringer}.
Further, exploring the entanglement in other types of mixed states, such as tripartite ground states of topological~\cite{Phys.Rev.B2018Shiozaki} and gapless systems~\cite{Phys.Rev.B2016Eisler}, and measurement-induced mixed states~\cite{Phys.Rev.Lett.2023Zhu} presents an intriguing avenue for further research.
Next, exploring the finite-size scaling laws of negativity in interacting fermionic systems could also be intriguing. 
In particular, our finding of the $L\ln L$ scaling of the RNR at the critical point in the spinless $t$-$V$ model may indicate long-range entanglement contribution~\cite{Phys.Rev.Lett.2020Wu,Phys.Rev.Res.2024Parez}, which warrants further investigation. 
Moreover, our results are applicable to the continuous-time QMC method, offering an opportunity to study the mixed-state entanglement of realistic correlated materials through combining with dynamical mean-field theory~\cite{Rev.Mod.Phys.2011Gull,Rev.Mod.Phys.1996Georges,Mod.Phys.Lett.B2013Su,Phys.Rev.Lett.2019Walsh,Phys.Rev.B2024Bellomia}. 
In the hybridization expansion algorithm, the bath can be firstly traced out~\cite{Phys.Rev.B2004Cheong,Phys.Rev.Lett.2013Grover}, allowing the impurity's reduced density matrix to be derived from the Green's functions and density correlation functions~\cite{Phys.Rev.Res.2024Parez}. 
Additionally, the rank-two RN can be computed within the interaction expansion algorithm via the identity ${\rm Tr}[(\rho^{T_{2}^{f}})^2] = {\rm Tr}[(\rho X_{2}(\pi))^2]$. Here, the disorder operator $X_{2}(\tau) \propto \prod_{j\in A_{2},\sigma} (n_{i\sigma}(\tau)-\frac{1}{2})$ in the interaction picture introduces additional interaction vertices exclusively in the $A_2$ subregion.
Finally, another proposal for realistic materials is to integrate the RN into the constrained-path auxiliary-field QMC method~\cite{Phys.Rev.Lett.1995Zhang,Phys.Rev.B1997Zhang,Phys.Rev.Lett.1999Zhang,Phys.Rev.B2019He} which controls the sign problem. 

\section{Methods}

\subsection{Determinantal Quantum Monte Carlo}
We used determinantal qauntum Monte Carlo (QMC) to simulate the two interacting fermionic models, both in zero-temperature regime (i.e., projective QMC) and in finite-temperature regime. Interesting readers may refer to the supplementary materials for all the details including basic formalism and the Hubbard-Stratonovich transformations used in this work. The projective DQMC calculations performed in Fig.~\ref{fig:HubbardChain} used a projection length $\Theta/t=20$, which is long enough to project the trial state to the ground state and ensures desired convergence. We chose the time slice step $\Delta_\tau$ to be between $0.02$ and $0.05$, depending on the size of $\Theta$ or $\beta$, and the results do not change if we choose a smaller $\Delta_\tau$. The results shown in Fig.~\ref{fig:HubbardChain} of the main text were accelerated by employing the delay update algorithm~\cite{Phys.Rev.B2024Sun}. 

\subsection{Incremental algorithm for R\'{e}nyi negativity}
The results presented in Fig.~\ref{fig:Negaraio-tV} were calculated using the incremental algorithm, which was proposed and implemented for R\'{e}nyi entanglement entropy~\cite{ArXiv2023Liaoa}. We have developed an analogous version for R\'{e}nyi negativity. Specifically, we measure the exponentially small observable $e^{-(n-1)\mathcal{E}_{n}}$ by separately calculating its $N_{\rm inc}$ factors, each of which is of order $O(10^{-1})$, 
\begin{equation}
    \begin{aligned}
    e^{-(n-1)\mathcal{E}_{n}}&=\frac{\sum_{\mathbf{s}}w_{\mathbf{s}_{1}}\cdots w_{\mathbf{s}_{n}}\det g_{x}}{\sum_{\mathbf{s}}w_{\mathbf{s}_{1}}\cdots w_{\mathbf{s}_{n}}}\\&=\frac{Z_{N_{{\rm inc}}}}{Z_{N_{{\rm inc}}-1}}\cdots\frac{Z_{k+1}}{Z_{k}}\cdots\frac{Z_{1}}{Z_{0}}, 
    \end{aligned}
\end{equation}
where $w_\mathbf{s}$ is the regular DQMC weight for a specific auxiliary field configuration $\mathbf{s}$. Here, we define intermediate partition functions as $Z_{k}=\sum_{{\bf s}_{1}\cdots{\bf s}_{n}}w_{{\bf s}_{1}}\cdots w_{{\bf s}_{n}}\left(\det g_{x}\right)^{\frac{k}{N_{\rm inc}}}$. Each ratio can be interpreted as the average value of the $N_{\rm inc}$-th root of the Grover determinant, $(\det g_x)^{1/N_{\rm inc}}$, over a replicated system with weight $W_{\mathbf{s}_1\cdots \mathbf{s}_n}=w_{{\bf s}_{1}}\cdots w_{{\bf s}_{n}}\left(\det g_{x}\right)^{\frac{k}{N_{\rm inc}}}$
\begin{equation}
    \begin{aligned}
    \frac{Z_{k+1}}{Z_{k}}&=\frac{\sum_{{\bf s}_{1}\cdots{\bf s}_{n}}w_{{\bf s}_{1}}\cdots w_{{\bf s}_{n}}\left(\det g_{x}\right)^{\frac{k+1}{N_{{\rm inc}}}}}{\sum_{{\bf s}_{1}\cdots{\bf s}_{n}}w_{{\bf s}_{1}}\cdots w_{{\bf s}_{n}}\left(\det g_{x}\right)^{\frac{k}{N_{{\rm inc}}}}}\\&=\frac{\sum_{{\bf s}_{1}\cdots{\bf s}_{n}}W_{{\bf s}_{1}\cdots{\bf s}_{n}}\left(\det g_{x}\right)^{\frac{1}{N_{{\rm inc}}}}}{\sum_{{\bf s}_{1}\cdots{\bf s}_{n}}W_{{\bf s}_{1}\cdots{\bf s}_{n}}}.
    \end{aligned}
\end{equation}
In our calculations, the number of intermediate processes is $N_{\rm inc}=64$, which is sufficiently large to achieve desirable statistical accuracy for the lattice sizes considered. 

\section{Data and code availability}

\paragraph{Data availability} The data that support the findings of this study are provided at \url{https://scidata.sjtu.edu.cn/records/2xm0w-eng25}.
\paragraph{Code availability} The code used in this work is available from the corresponding author upon request.

\section{Acknowledgements}
\begin{acknowledgments}
    The authors thank Tarun Grover for helpful discussions and comments on the draft.
    This work was supported by the National Key R\&D Program of China (Grant No. 2022YFA1402702, No. 2021YFA1401400), the National Natural Science Foundation of China (Grants No. 12447103, No. 12274289), the Innovation Program for Quantum Science and Technology (under Grant No. 2021ZD0301902), Yangyang Development Fund, and startup funds from SJTU. The computations in this paper were run on the Siyuan-1 and $\pi$ 2.0 clusters supported by the Center for High Performance Computing at Shanghai Jiao Tong University.
\end{acknowledgments}

\bibliography{reference_nega.bib}

\end{document}


\title{Supplementary Information for \\``Entanglement R\'{e}nyi negativity of interacting fermions from quantum Monte Carlo simulations''}
\author{Fo-Hong Wang}
\author{Xiao Yan Xu*}

\maketitle
\vspace{-3em}
\begin{center}
{\small *Corresponding author. Email: xiaoyanxu@sjtu.edu.cn}
\end{center}

This PDF file includes:
\begin{itemize}
    \item \supnameref{sec:FPT}
    \item \supnameref{sec:DQMC}
    \item \supnameref{sec:DQMC_FPT}
    \item \supnameref{sec:additional_results}
    \item \supnameref{sec:sign_problem}
    \item Supplementary Figs. \ref{fig:nega-entro1} and \ref{fig:nega-tvdata}
\end{itemize}

\setcounter{equation}{0}
\setcounter{figure}{0}
\setcounter{table}{0}
\setcounter{page}{1}
\makeatletter

\section{Fermionic partial transpose in different representations}\label{sec:FPT}

In this section, we briefly review the definition of fermionic partial transpose, which does not follow the original definition~\cite{Phys.Rev.A2002Vidal} extensively used in bosonic systems. 
Consider a lattice model described by complex fermion operators $c_j$ and $c^\dagger_j$ satisfying anticommunication relations $\{c_j,c^\dagger_k\}=\delta_{jk}$, where $j,k=1,\dots,N$ are labels of sites (for simplicity we omit the index for internal degree of freedom). For convenience, we also introduce the Majorana basis, denoted as $\gamma_{2j-1}=c_{j}+c_{j}^{\dagger}$ and $\gamma_{2j}=-\text{i}(c_{j}-c_{j}^{\dagger})$. Under a bipartite scheme that divides the total system as $A=A_1\cup A_{2}$, the fermionic partial transpose with respect to subsystem $A_2$, denoted by $\mathcal{O}^{T_2^f}$ with $\mathcal{O}$ as an operator (like the density operator $\rho$ or just a single basis operator $|\{e_j\}\rangle\langle \{\bar{e}_j\} |$), is first defined in the coherent basis as~\cite{Phys.Rev.B2018Shiozaki}
\begin{equation}\label{equ:FPT_coherent}
    \begin{aligned}
        U_{A_{2}}\left(|\{\xi_{j}\}_{j\in A_{1}},\{\xi_{j}\}_{j\in A_{2}}\rangle\langle\{\bar{\chi}_{j}\}_{j\in A_{1}},\{\bar{\chi}_{j}\}_{j\in A_{2}}|\right)^{T_{2}^{f}}U_{A_{2}}^{\dagger}=|\{\xi_{j}\}_{j\in A_{1}},\{-\text{i}\bar{\chi}_{j}\}_{j\in A_{2}}\rangle\langle\{\bar{\chi}_{j}\}_{j\in A_{1}},\{-\text{i}\xi_{j}\}_{j\in A_{2}}|,
    \end{aligned}
\end{equation}
where $|\{\xi_{j}\}\rangle=e^{-\sum_{j}\xi_{j}c_{j}^{\dagger}}|0\rangle$ and $\langle\{\bar{\chi}_{j}\}|=\langle0|e^{-\sum_{j}c_{j}\bar{\chi}_{j}}$ are the fermion coherent states, and $U_{A_2}\equiv \prod_{j\in A_2}\gamma_{2j-1}$ is the partial particle-hole transformation which only turns the particles (holes) in the subsystem $A_2$ into holes (particles). After choosing an appropriate ordering such that $A_2=\{N_{1}+1,\dots,N\}$, one obtains fermionic partial transpose in the occupation number basis by expressing the coherent states in Supplementary Eq.~\eqref{equ:FPT_coherent} in terms of Fock states~\cite{Phys.Rev.B2018Shiozaki,J.Stat.Mech.2019Shapourian}
\begin{equation}\label{equ:FPT_occup}
    \begin{aligned}
        (|\{ n_{j}\} _{j\in A_{1}},\{ n_{j}\} _{j\in A_{2}}\rangle \langle \{ \bar{n}_{j}\} _{j\in A_{1}},\{ \bar{n}_{j}\} _{j\in A_{2}}|)^{T_{2}^{f}}=(-1)^{\phi(\{ n_{j}\} ,\{ \bar{n}_{j}\} )}
        |\{ n_{j}\} _{j\in A_{1}},\{ \bar{n}_{j}\} _{j\in A_{2}}\rangle \langle \{ \bar{n}_{j}\} _{j\in A_{1}},\{ n_{j}\} _{j\in A_{2}}|,
    \end{aligned}
\end{equation}
which is similar to the conventional partial transpose up to an additional phase factor 
\begin{equation}
  \phi\left(\left\{ n_{j}\right\} ,\left\{ \bar{n}_{j}\right\} \right)=\left[\left(\tau_{2}+\bar{\tau}_{2}\right)\bmod2\right]/2+\left(\tau_{1}+\bar{\tau}_{1}\right)\left(\tau_{2}+\bar{\tau}_{2}\right)
\end{equation}
with $\tau_b=\sum_{j \in A_b} n_j$ the number of particles in subsystem $A_b$ ($b=1,2$). The definition in Supplementary Eq.~\eqref{equ:FPT_occup} has been employed in the exact diagonalization calculations in Fig.~1 of the main text. 
Moreover, utilizing this definition, it is straightforward to show that
\begin{equation}
    \begin{aligned}
        \mathrm{Tr}\left[\left(\rho^{T_{2}^{f}}\right)^{k}\right]=&\sum_{\left\{ n_{j}^{\left(1\right)}\right\} \dots\left\{ n_{j}^{\left(k\right)}\right\} }\left\langle \left\{ n_{j\in A_{1}}^{\left(1\right)},n_{j\in A_{2}}^{\left(3\right)}\right\} \middle|\rho\middle|\left\{ n_{j}^{\left(2\right)}\right\} \right\rangle \left\langle \left\{ n_{j\in A_{1}}^{\left(2\right)},n_{j\in A_{2}}^{\left(4\right)}\right\} \middle|\rho\middle|\left\{ n_{j}^{\left(3\right)}\right\} \right\rangle \cdots\left\langle \left\{ n_{j\in A_{1}}^{\left(k\right)},n_{j\in A_{2}}^{\left(2\right)}\right\} \middle|\rho\middle|\left\{ n_{j}^{\left(1\right)}\right\} \right\rangle \\&\times\left(-1\right)^{\phi\left(\left\{ n_{j}^{\left(1\right)}\right\} ,\left\{ n_{j}^{\left(2\right)}\right\} \right)}\left(-1\right)^{\phi\left(\left\{ n_{j}^{\left(2\right)}\right\} ,\left\{ n_{j}^{\left(3\right)}\right\} \right)}\cdots\left(-1\right)^{\phi\left(\left\{ n_{j}^{\left(k\right)}\right\} ,\left\{ n_{j}^{\left(1\right)}\right\} \right)}.
    \end{aligned}
\end{equation}
Specifically, note that $(-1)^{\phi(\{n_j\},\{n_j\})}=1$ and $\phi(\{n_j\},\{\bar{n}_j\})=\phi(\{\bar{n}_j\},\{n_j\})$, one can deduce that $\mathrm{Tr}[\rho^{T_2^f}]=\mathrm{Tr}[\rho]=1$ while ${\rm Tr} [ (\rho^{T_{2}^{f}} )^{2} ]={\rm Tr}[ (\rho\hat{X}_{2} (\pi ))^2]$ with $\hat{X}_{2}\left(\theta\right)=e^{{\rm i}\theta\sum_{j\in A_{2}}\hat{n}_{j}}$ being the disorder operator and $\hat{n}_j=c^\dagger_j c_j$ the local density operator~\cite{JStatPhys2017Fradkin,Phys.Rev.Lett.2024Liua}.

In general, density operators can be written as a restricted superposition of products of Majorana operators. Assume that there are $k$ ($l$) sites in subsystem $A_1$ ($A_2$), in which the Majorana indices are denoted by $\{m_{1},\dots,m_{2k}\}$ ($\{n_{1},\dots,n_{2l}\}$), a density operator can be expressed as~\cite{NewJ.Phys.2015Eisler,Phys.Rev.B2017Shapourian}
\begin{equation}\label{equ:rho_Majorana}
    \rho=\sum_{\substack{\kappa,\tau,\\|\kappa|+|\tau|=\text{even}}}w_{\kappa,\tau}\gamma_{m_{1}}^{\kappa_{1}}\cdots\gamma_{m_{2k}}^{\kappa_{2k}}\gamma_{n_{1}}^{\tau_{1}}\cdots\gamma_{n_{2l}}^{\tau_{2l}}
\end{equation}
where $\kappa=(\kappa_{1},\cdots,\kappa_{2k})$ and $\tau=(\tau_{1},\cdots,\tau_{2l})$ represent various Majorana configurations. Here, $\kappa_{i}$ and $\tau_{j}$ are the occupations of single Majorana modes, and $|\kappa|=\sum_{j}\kappa_{j}$ or $|\tau|=\sum_{j}\tau_{j}$ is the total number of Majorana fermions in the corresponding subsystem. We note that $w_{\kappa,\tau}\neq0$ only if $|\kappa|+|\tau|$ is even since a physical state must have a specific parity. Now, we evaluate the fermionic partial transpose $\rho^{T_2^f}$ based on Supplementary Eq.~\eqref{equ:rho_Majorana}, and for each term, the operators in subsystem $A_2$ would be transformed to $\mathcal{R}_2^f(\gamma_{n_{1}}^{\tau_{1}}\cdots\gamma_{n_{2l}}^{\tau_{2l}})$. It turns out that the definition in Supplementary Eq.~\eqref{equ:FPT_coherent} would give us a simple expression for the transformation $\mathcal{R}_2^f$~\cite{Phys.Rev.B2017Shapourian},
\begin{equation}\label{equ:FPT_majorana_SM}
    \mathcal{R}_2^f(\gamma_j)=\text{i}\gamma_j,\quad j\in A_2.
\end{equation}
Under this fermionic partial transpose, a Gaussian state $\rho_0=\det[1+e^W]^{-1/2}\exp\left(\frac{1}{4}\sum_{k,l}W_{kl}\gamma_{k}\gamma_{l}\right)$ will be transformed to another Gaussian state. 

\section{Determinant Quantum Monte Carlo Methods}\label{sec:DQMC}

In this section, we provide a brief introduction to determinant quantum Monte Carlo (DQMC) methods~\cite{Phys.Rev.D1981Blankenbecler,Phys.Rev.B1981Scalapino,Assaad2008WorldlineDeterminantalQuantum}. For our purpose, both the zero-temperature projector scheme and the finite-temperature scheme have been used in the main text. 

\subsection{Finite-temperature Scheme}

At a finite temperature $T$, and assuming that the system of interest is in thermodynamic equilibrium, we can analyze it within the framework of the grand canonical ensemble, using the partition function $Z=\operatorname{Tr}\left[e^{-\beta H}\right]$. A generic Hamiltonian $H$ consists of a free-particle term and an interaction term, denoted as $H=H_0+H_I$. To compute the trace over Fock space, we employ Trotter decomposition and Hubbard-Stratonovich (HS) transformation to factorize the exponential operator $e^{-\beta H}$ into a sum of products of Gaussian operators, 
\begin{equation}\label{equ:S2_partition}
    \begin{aligned}
        Z&=\operatorname{Tr}\left[e^{-\beta H}\right]=\operatorname{Tr}\left[\left(e^{-\Delta_{\tau}H}\right)^{L_{\tau}}\right]\\&=\operatorname{Tr}\left[e^{-\Delta_{\tau}H_{I}}e^{-\Delta_{\tau}H_{0}}\cdots e^{-\Delta_{\tau}H_{I}}e^{-\Delta_{\tau}H_{0}}\right]+O\left(\Delta_{\tau}^{2}\right)\\&\approx\sum_{\left\{ s_{i,l}\right\} }\operatorname{Tr}\left[\prod_{l=1}^{L_{\tau}}\left(e^{{\bf c}^{\dagger}V\left({l}\right){\bf c}}e^{\bm{c}^{\dagger}K{\bf c}}\right)\right],
    \end{aligned}
\end{equation}
where $L_\tau=\beta/\Delta_\tau$ is the number of time slices, $-\Delta_{\tau}H_{0}={\bf c^{\dagger}}K{\bf c}$ with $\mathbf{c}=(c_1,\dots,c_N)^T$, and we have decoupled the interaction term $H_I$ to fermion bilinears ${\bf c}^{\dagger}V(l){\bf c}={\bf c}^{\dagger}V[\mathbf{s}(l)]{\bf c}$ coupled with spacetime-dependent auxiliary fields $\mathbf{s}=\{s_{i,l},i\in 1,\dots,N_c;l=1,\dots,L_\tau\}$. Here, $\mathbf{s}(l)$ includes all the auxiliary fields at time slice $l$ and $N_c$ represents the number of coupling terms, which varies depending on the specific interactions and decoupled channels. 
For the Hubbard model, we decouple it to the density channel, 
\begin{equation}\label{S2_HSHubbard}
    e^{-\Delta_{\tau}\frac{U}{2}\sum_{i}\left(n_{i}-1\right)^{2}}=\sum_{\left\{ s_{i}=\pm1,\pm2\right\} }\left(\prod_{i}\gamma(s_{i})e^{-{\rm i}\sqrt{\Delta_{\tau}U/2}\eta(s_{i})}\right)e^{{\rm i}\sqrt{\Delta_{\tau}U/2}\sum_{i}\eta(s_{i})n_{i}},
\end{equation}
where $ \gamma( \pm 1)=1+\sqrt{6} / 3, \gamma( \pm 2)=1-\sqrt{6} / 3, \eta( \pm 1)= \pm \sqrt{2(3-\sqrt{6})}$ and $ \eta( \pm 2)= \pm \sqrt{2(3+\sqrt{6})}$. Thus, for the Hubbard model, $N_c$ is the number of sites $N$. For the spinless $t$-$V$ model, we decouple it to the hopping channel~\cite{Phys.Rev.B2015Li,Phys.Rev.Lett.2015Wang},
\begin{equation}\label{S2_HStV}
    e^{-\Delta\tau V\sum_{\left\langle jk\right\rangle }\left(n_{j}-\frac{1}{2}\right)\left(n_{k}-\frac{1}{2}\right)}=\sum_{\left\{ s_{jk}=\pm1\right\} }\left(\frac{1}{2}e^{-\frac{V\Delta\tau}{4}}\right)e^{\lambda\sum_{\left\langle jk\right\rangle }s_{jk}\left(c_{j}^{\dagger}c_{k}+c_{k}^{\dagger}c_{j}\right)},
\end{equation}
where $\cosh \lambda=e^{\frac{V \Delta \tau}{2}}$. Thus, for the $t$-$V$ model, the subscript $i$ of auxiliary fields ${s}_{i,l}$ denotes various nearest neighboring (NN) bonds $\langle jk\rangle$, and $N_c$ represents the number of NN bonds (specifically, for bipartite lattices $N_c=Nz/2$ with $z$ being the coordination number). The trace of products of Gaussian operators over the fermionic Fock space in the last line of Supplementary Eq.~\eqref{equ:S2_partition} can be expressed as a determinant,
\begin{equation}\label{equ:S2_partition2}
    Z=\sum_{\mathbf{s}}\omega_\mathbf{s}=\sum_{\mathbf{s}}\alpha_\mathbf{s}\det\left[I+\prod_{l=L_{\tau}, \cdots, 1} B_\mathbf{s}(l)\right]\text{ with }B_\mathbf{s}(l)=e^{{V}(l)} e^{{K}},
\end{equation}
where $\alpha_\mathbf{s}$ denotes the coefficients before the exponential operators. 
For the Hubbard model, $\alpha_\mathbf{s}=(\prod_{i,l}\gamma(s_{i,l})e^{-{\rm i}\sqrt{\Delta_{\tau}U/2}\eta(s_{i,l})})$, while for the $t$-$V$ model, $\alpha_\mathbf{s}=\prod_{\langle jk\rangle,l}(\frac{1}{2}e^{-\frac{V\Delta\tau}{4}})$.
In addition, the expectation of arbitrary operator $O$ can also be decomposed into a sum over auxiliary fields, 
\begin{equation}\label{equ:S2_expctO}
    \left\langle O\right\rangle =\frac{\operatorname{Tr}\left[\mathrm{e}^{-\beta H}O\right]}{\operatorname{Tr}\left[\mathrm{e}^{-\beta H}\right]}=\sum_{{\bf s}}P_{{\bf s}}\langle O\rangle_{{\bf s}}+O\left(\Delta_{\tau}^{2}\right)\text{ with }P_\mathbf{s}=\frac{\omega_\mathbf{s}}{\sum_\mathbf{s}\omega_\mathbf{s}}.
\end{equation}
Here, the expectation of $O$ with respect to a specific configuration of auxiliary field is given by
\begin{equation}
    \langle O\rangle_{\mathbf{s}}=\frac{\operatorname{Tr}\left[U_{\mathbf{s}}(\beta,\tau)OU_{\mathbf{s}}(\tau,0)\right]}{\operatorname{Tr}U_{\mathbf{s}}(\beta,0)}\text{ with }U_{\mathbf{s}}(\tau_{2}=l_{2}\Delta_{\tau},\tau_{1}=l_{1}\Delta_{\tau})=\prod_{l=l_{1}+1}^{l_{2}}\left(e^{{\bf c}^{\dagger}{V}(l){\bf c}}e^{{\bf c}^{\dagger}{K}{\bf c}}\right).
\end{equation}
For instance, the most elementary observable, namely the equal-time Green's function, can be calculated via $G_{{\bf s},ij}(\tau,\tau)=\langle c_{i}c_{j}^{\dagger}\rangle _{{\bf s}}=(1+B_{{\bf s}}(\tau,0)B_{{\bf s}}(\beta,\tau))_{ij}^{-1}$ where $B_{\mathbf{s}}(\tau_{2}=l_{2}\Delta_{\tau},\tau_{1}=l_{1}\Delta_{\tau})=\prod_{l=l_{1}+1}^{l_{2}}\left(e^{{V}(l)}e^{{K}}\right)$ is the matrix correspondence of $U_\mathbf{s}(\tau_2,\tau_1)$. 

In practice, the weighted sums in Supplementary Eqs.~\eqref{equ:S2_partition2} and \eqref{equ:S2_expctO} are conducted by using Monte-Carlo importance sampling. 
Specifically, we adopt a local update scheme where the auxiliary fields are updated sequentially, and the acceptance ratio is determined by the Metropolis algorithm. 
Let us consider an update which is proposed to happen at time slice $l$ and site $i$, i.e., $s_{i,l}\rightarrow s_{i,l}^\prime$, which leads to a change of the determinant weight $\det[I+\prod_l B_\mathbf{s}(l)]$: $e^{V^{\prime}(l)}\equiv(I+\Delta) e^{V(l)}$. 
For simplicity, we assume that $\Delta$ is a diagonal matrix with $k$ nonzero elements. 
Whether the update is accepted or not is determined by the Metropolis acceptance probability $P_{\text{acc}}=\min\left\{1,\frac{\omega_{\mathbf{s}^\prime}}{\omega_{\mathbf{s}}}\right\}$, where the update ratio can be calculated as
\begin{equation}\label{equ:S2_update_ratio}
    \begin{aligned}
        \frac{\omega_{{\bf s}^{\prime}}}{\omega_{{\bf s}}}&=\frac{\alpha_{{\bf s}^{\prime}}}{\alpha_{{\bf s}}}\frac{\det\left[I+B_{\mathbf{s}}\left(\beta,\tau\right)\left(I+\Delta\right)B_{\mathbf{s}}\left(\tau,0\right)\right]}{\det\left[I+B_{\mathbf{s}}\left(\beta,0\right)\right]}\\&=\frac{\alpha_{{\bf s}^{\prime}}}{\alpha_{{\bf s}}}\frac{\det G_{\mathbf{s}}\left(\tau,\tau\right)}{\det G_{\mathbf{s}^{\prime}}\left(\tau,\tau\right)}=\frac{\alpha_{{\bf s}^{\prime}}}{\alpha_{{\bf s}}}\det\left[I_{k}+VU\right]
    \end{aligned}
\end{equation}
Here, we have used the fast update formula for the Green's function matrix, 
\begin{equation}
    G_{\mathbf{s}^{\prime}}\left(\tau,\tau\right)=G_{\mathbf{s}}\left(\tau,\tau\right)-G_{\mathbf{s}}\left(\tau,\tau\right)U(I_{k}+VU)^{-1}V, 
\end{equation}
where we define $U\equiv P_{N\times k}D$ and $V\equiv P_{k\times N}(I-G_{\mathbf{s}_i}(\tau,\tau))$. They are rectangular matrices that satisfy $\Delta(I-G_\mathbf{s}) \equiv U V$. 
The projection matrices $P_{N\times k}$ and $P_{k\times N}$ are cropped from the identity matrix $I=I_{N}$, with columns or rows corresponding to the non-zero entries of $\Delta$. 
If the update is accepted, then we use the above formula to update $G_\mathbf{s}$. 
After finishing the updates of all sites at time slice $l$, we move to the next time slice by propagating the Green's function matrix using
\begin{equation}
    \begin{aligned}
        G_{\mathbf{s}}\left(\tau+\Delta_{\tau},\tau+\Delta_{\tau}\right)&=B_{\mathbf{s}}\left(l+1\right)G_{\mathbf{s}}(\tau,\tau)B_{\mathbf{s}}\left(l+1\right)^{-1},\\G_{\mathbf{s}}\left(\tau-\Delta_{\tau},\tau-\Delta_{\tau}\right)&=B_{\mathbf{s}}\left(l\right)^{-1}G_{\mathbf{s}}(\tau,\tau)B_{\mathbf{s}}\left(l\right).
        \end{aligned}
\end{equation}
At each time slice, we perform one measurement of the observables of interest, which can be expressed by the Green's functions by means of Wick's theorem. 
For example, the density-density correlation function can be calculated as $\langle n_i n_j\rangle_\mathbf{s}=\langle c_i^\dagger c_i\rangle_\mathbf{s} \langle c_j^\dagger c_j\rangle_\mathbf{s}+\langle c_i^\dagger c_j\rangle_\mathbf{s} \langle c_i c_j^\dagger\rangle_\mathbf{s}$. 
After several updates across different time slices, the Green's function matrix requires numerical stabilization based on matrix decomposition. 
For instance, to stably calculate $G_\mathbf{s}(\tau,\tau)$, we can first perform singular value decompositions: $B_\mathbf{s}(\beta,\tau)^\dagger=U_LD_LV_L$ and $B_\mathbf{s}(\tau,0)=U_RD_RV_R$, and then calculate $G_\mathbf{s}(\tau,\tau)$ using
\begin{equation}
    G_{\mathbf{s}}\left(\tau,\tau\right)=U_{L}D_{L+}^{-1}\left(D_{R+}^{-1}U_{R}^{\dagger}U_{L}D_{L+}^{-1}+D_{R-}V_{R}V_{L}^{\dagger}D_{L-}\right)^{-1}D_{R+}^{-1}U_{R}^{\dagger},
\end{equation}
where $D_{L/R,+}=\max (D_{L/R}, 1)$ and $D_{L/R,-}=\min (D_{L/R}, 1)$. 

\subsection{Zero-temperature Projector Scheme}

In the projector DQMC scheme, the ground-state wavefunction of interest is calculated by projecting from a trial wavefunction $|\Psi_T\rangle$. It is important to note that the trial wavefunction should not be orthogonal to the true ground state $|\psi_0\rangle$ so that it is possible to obtain $|\psi_0\rangle=\lim_{\Theta\rightarrow \infty}e^{-\Theta H}|\Psi_T\rangle$ for a sufficiently long projection length $\Theta$. Analogous to the finite temperature case, the modulus of the ground state (which plays the role of the ``partition function'') and the ground-state expectation of some observable ${O}$ are written as summations over auxiliary fields after doing Trotter decomposition and HS transformation. The magnitude of the ground state is given by
\begin{equation}\label{equ:S2_partition_proj}
    \left\langle \Psi_{0}\mid\Psi_{0}\right\rangle =\left\langle \Psi_{T}\left|e^{-2\Theta H}\right|\Psi_{T}\right\rangle \approx\sum_{{\bf s}}\omega_{{\bf s}}=\sum_{{\bf s}}\alpha_\mathbf{s}\det\left[{P}^{\dagger}{B}_\mathbf{s}(2\Theta,0){P}\right],
\end{equation}
where $P$ is the coefficient matrix of the trial state determined by $|\Psi_{T}\rangle=\prod_{n=1}^{N_{p}}\left({\bf c}^{\dagger}P\right)_{{n}}|0\rangle$, with $N_p$ representing the number of occupied single-particle states. Here one can also see that $2\Theta$ plays a similar role to $\beta$ in the finite-temperature case. The observable expectation is given by 
\begin{equation}\label{equ:S2_expctO_proj}
    \langle{O}\rangle=\sum_{\mathbf{s}}{P}_{\mathbf{s}}\langle{O}\rangle_{\mathbf{s}}=\frac{\sum_{\mathbf{s}}{\omega}_{\mathbf{s}}\langle{O}\rangle_{\mathbf{s}}}{\sum_\mathbf{s}\omega_\mathbf{s}}\text{ with }
    \langle O\rangle_{{\bf s}}=\frac{\left\langle \Psi_{T}\left|U_{{\bf s}}\left(2\Theta,\tau\right)OU_{{\bf s}}\left(\tau,0\right)\right|\Psi_{T}\right\rangle }{\left\langle \Psi_{T}\left|e^{-2\Theta H}\right|\Psi_{T}\right\rangle }.
\end{equation}
For instance, the equal-time Green's function can be calculated as ${G}_\mathbf{s}(\tau, \tau)=I-{R}(\tau)({L}(\tau) {R}(\tau))^{-1} {L}(\tau)$ with ${R}(\tau)={B}_\mathbf{s}(\tau, 0) {P}$ an $N\times N_p$ matrix, and ${L}(\tau)={P}^{\dagger} {B}_\mathbf{s}(2 \Theta, \tau)$ an $N_p\times N$ matrix. 

In practice, we use a similar Monte-Carlo importance sampling method as the finite-temperature case to evaluate the weighted sums in Supplementary Eqs.~\eqref{equ:S2_partition_proj} and \eqref{equ:S2_expctO_proj}. 
Instead of updating the Green's function matrix $G$, it is more efficient to update the matrices $R$ and $(LR)^{-1}$ using
\begin{equation}
    R^\prime=(I+\Delta)R\text{ and }\left[\left(LR\right)^{-1}\right]^{\prime}=(LR)^{-1}-(LR)^{-1}U(I_{k}+VU)^{-1}V, 
\end{equation}
where $U\equiv L\Delta P_{N\times k}$ and $V\equiv P_{k\times N}R(LR)^{-1}$. 
The acceptance ratio is given by
\begin{equation}
    \frac{\omega_{\mathbf{s}^{\prime}}}{\omega_{\mathbf{s}}}=\frac{\alpha_{\mathbf{s}^{\prime}}\det\left[\left(LR\right)^{\prime}\right]}{\alpha_{\mathbf{s}}\det\left[LR\right]}=\frac{\alpha_{\mathbf{s}^{\prime}}}{\alpha_{\mathbf{s}}}\det\left[I_{k}+VU\right].
\end{equation} 
The propagation of $L(\tau)$ and $R(\tau)$ is straightforward using
\begin{equation}
    L(\tau+\Delta_\tau)=L(\tau)B_\mathbf{s}(l+1)^{-1} \text{ and } R(\tau+\Delta_\tau)=B_\mathbf{s}(l+1)R(\tau).
\end{equation}
As for the measurements, we note that to obtain an accurate representation of the true ground state, it is advisable to only perform measurements around $\tau=\Theta$. 
Finally, the numerical stabilization of Green's function matrix is much easier in the projector scheme. 
For example, we can first perform QR decompositions: $L_{N_{p} \times N}=D_{L}\left(U_{L}\right)_{N_{p} \times N}$ and $R_{N \times N_{p}}=\left(U_{R}\right)_{N \times N_{p}} D_{R}$, and then calculate $G(\Theta,\Theta)$ using
\begin{equation}
    G=1-R(LR)^{-1}L=1-U_{R}\left(U_{L}U_{R}\right)^{-1}U_{L}.
\end{equation}
According to the above formula, we can update the $U_R$, $U_L$ and $(U_LU_R)^{-1}$ matrices instead of $R$, $L$ and $(LR)^{-1}$, respectively, which is more stable and utilized in practice. 


\section{DQMC implementation of fermionic partial transpose}\label{sec:DQMC_FPT}

In this section, we provide a comprehensive discussion of the DQMC implementation of fermionic partial transpose. For completeness, we begin with the formulation of DQMC in the Majorana basis and then transition back to the complex fermion basis.

\subsection{Decomposition of interacting density matrix}

Before starting the discussion, we first introduce the so-called \textit{product rule of Gaussian states}, which means that the product of two Gaussian states is still a Gaussian state. 
This property can be proved using the Baker-Campbell-Hausdorff formula,
\begin{equation}
    \begin{aligned}
    e^{\hat{W}^{\prime\prime}}&\equiv e^{\hat{W}}e^{\hat{W}^{\prime}}
    =e^{\hat{W}+\hat{W}^{\prime}+\frac{1}{2}\left[\hat{W},\hat{W}^{\prime}\right]+\frac{1}{12}\left(\left[\hat{W},\left[\hat{W},\hat{W}^{\prime}\right]\right]+\left[\hat{W}^{\prime},\left[\hat{W},\hat{W}^{\prime}\right]\right]\right)+\cdots},
    \end{aligned}
\end{equation}
and the following commuting relations
\begin{subequations}
    \begin{equation}
        \left[\gamma_{l}\gamma_{n},\gamma_{j}\gamma_{k}\right]=-2\gamma_{l}\gamma_{j}\delta_{kn}+2\gamma_{l}\gamma_{k}\delta_{jn}+2\gamma_{n}\gamma_{j}\delta_{lk}-2\gamma_{n}\gamma_{k}\delta_{lj},
    \end{equation}
    \begin{equation}
        \left[c_{l}^{\dagger}c_{n},c_{j}^{\dagger}c_{k}\right]=c_{l}^{\dagger}c_{k}\delta_{nk}-c_{j}^{\dagger}c_{n}\delta_{lk}.
    \end{equation}
\end{subequations}
As a result, provided that both $e^{\hat{W}}$ and $e^{\hat{W}^{\prime}}$ are Gaussian states (up to normalization), we can assert that $\hat{W}^{\prime\prime}$ is a fermionic quadratic, thus $e^{\hat{W}^{\prime\prime}}$ is a Gaussian state (up to normalization). 

In the framework of DQMC, the partition function is given by $Z=\sum_{\mathbf{s}}\mathrm{Tr}[\prod_{l=1}^{L_\tau} e^{\mathbf{c}^\dagger K_{l}[\mathbf{s}]\mathbf{c}}]$, where $K_l$ combines the $K$ and $V(l)$ in Supplementary Eq.~\eqref{equ:S2_partition} by means of the product rule of Gaussian states. 
We then convert to the Majorana basis by rewriting the decoupled Hamiltonian as $\mathbf{c}^\dagger K_l[\mathbf{s}]\mathbf{c}=\boldsymbol{\gamma}^T h_l[\mathbf{s}]\boldsymbol{\gamma}/4$, where $h_l$ is a $2N\times 2N$ antisymmetric matrix satisfying the antisymmetric condition $h_l=-h_l^T$ and $\bm{\gamma}=(\gamma_1,\dots,\gamma_{2N})$. 
It is worth noting that this form can encompass terms beyond particle-number conserving terms, such as pairing terms~\cite{Phys.Rev.C1993Lang,J.Stat.Mech.2014Klich,ArXiv2018Wei}. The partition function in the Majorana basis is given by~\cite{J.Stat.Mech.2014Klich,Phys.Rev.B2015Li,ArXiv2018Wei}
\begin{equation}\label{equ:S3_partition}
    \begin{aligned}
        Z&=\sum_{{\bf s}}\text{Tr}\left[e^{\frac{1}{4}\bm{\gamma}^{T}h_{L_{\tau}}[{\bf s}]\bm{\gamma}}\cdots e^{\frac{1}{4}\bm{\gamma}^{T}h_{l}[{\bf s}]\bm{\gamma}}\cdots e^{\frac{1}{4}\bm{\gamma}^{T}h_{1}[{\bf s}]\bm{\gamma}}\right]=\sum_{\mathbf{s}}\text{Tr}\left[e^{\frac{1}{4}\bm{\gamma}^{T}W_{\mathbf{s}}\bm{\gamma}}\right]\\&=\sum_{{\bf s}}\pm\det\left[I+e^{h_{L_{\tau}}}\cdots e^{h_{l}}\cdots e^{h_{1}}\right]^{1/2}=\sum_{{\bf s}}\pm\det\left[I+e^{W_{\mathbf{s}}}\right]^{1/2},
        \end{aligned}
\end{equation}
where we have introduce $W_{\mathbf{s}}$ by further utilizing the product rule of Gaussian states. 
Therefore, the density matrix of an interacting system can be decomposed to a sum of Gaussian states~\cite{Phys.Rev.Lett.2013Grover}
\begin{equation}\label{equ:S3_rho_decomp}
    \rho=\sum_{{\bf s}}p_{{\bf s}}\rho_{{\bf s}},\quad \rho_{\bm{s}}=\det\left[I+e^{{W}_\mathbf{s}}\right]^{-1/2}e^{\frac{1}{4}\boldsymbol\gamma^{T}W_\mathbf{s}\boldsymbol\gamma}. 
\end{equation}

\subsection{Fermionic partial transpose of a single Gaussian state}

First of all, we discuss the fermionic partial transpose of a single Gaussian state, i.e. the $\rho_\mathbf{s}$ in Supplementary Eq.~\eqref{equ:S3_rho_decomp}.
We may abbreviate the subscript $\mathbf{s}$. These results are related to the Supplementary Eqs.~(2) and (3), and also the statements around them in the main text. 

The ``Green's function'' in the Majorana basis is defined as $\Gamma_{\mathbf{s},kl}=\langle[\gamma_{k},\gamma_{l}]\rangle_{{\bf s}}/2=\langle\frac{1}{4}\gamma^{T}O^{kl}\gamma\rangle_{{\bf s}}$ with matrix $O_{ij}^{kl}=2\delta_{ik}\delta_{jl}-2\delta_{il}\delta_{jk}$. $\Gamma$ is also called covariance matrix which characterizes a Gaussian state with relation $\tanh(-W/2)=\Gamma$~\cite{NewJ.Phys.2015Eisler} or inverse relation $W=\ln\left[(I+\Gamma)^{-1}(I-\Gamma)\right]$. 
To prove these relations, we treat $\frac{1}{4}\boldsymbol{\gamma}^TO^{kl}\boldsymbol{\gamma}$ as a generic observable and calculate its expectation as
\begin{equation}\label{equ:S3_Gamma1}
    \begin{aligned}
        \Gamma_{\mathbf{s},kl}&=\langle\frac{1}{4}\bm{\gamma}^{T}O^{kl}\bm{\gamma}\rangle_{\mathbf{s}}=\mathrm{Tr}\left[\rho_{\mathbf{s}}\frac{1}{4}\bm{\gamma}^{T}O^{kl}\bm{\gamma}\right]\\&=\left.\frac{\partial}{\partial\eta}\ln\text{Tr}\left[e^{\frac{1}{4}\bm{\gamma}^{T}W_{\mathbf{s}}\bm{\gamma}}e^{\frac{1}{4}\eta\bm{\gamma}^{T}O^{kl}\bm{\gamma}}\right]\right|_{\eta=0}=\frac{1}{2}\left.\frac{\partial}{\partial\eta}\text{Tr}\ln\left[I+e^{W_{\mathbf{s}}}e^{\eta O^{kl}}\right]\right|_{\eta=0}\\&=\frac{1}{2}\text{Tr}\left[(I+e^{W_{\mathbf{s}}})^{-1}e^{W_{\mathbf{s}}}O^{kl}\right]=\frac{1}{2}\left[(I+e^{W_{\mathbf{s}}})^{-1}e^{W_{\mathbf{s}}}\right]_{ji}O_{ij}^{kl}\\&=\left[(I+e^{W_{\mathbf{s}}})^{-1}e^{W_{\mathbf{s}}}\right]_{lk}-\left[(I+e^{W_{\mathbf{s}}})^{-1}e^{W_{\mathbf{s}}}\right]_{kl}. 
    \end{aligned}
\end{equation}
Based on the above relation, we can prove the relation between $\Gamma$ and $W$, 
\begin{equation}
    \begin{aligned}
        \Gamma&=-(I+e^{W})^{-1}e^{W}+e^{-W}(I+e^{-W})^{-1}\\&=-\left(e^{W/2}-e^{-W/2}\right)\left(e^{W/2}+e^{-W/2}\right)^{-1}=\tanh\left(-W/2\right).
    \end{aligned}
\end{equation}
This relation is of importance because it indicates that a covariance matrix can be used to describe a Gaussian state equivalently. 
We note that the above proof assumes that the trace in Supplementary Eq.~\eqref{equ:S3_partition} is without sign ambiguity. One can find a more general proof in Supplementary Ref.~\cite{J.Stat.Mech.2010Fagotti}.

The ``partial transpose of the Green's function'' in Majorana basis is quite straightforward via Supplementary Eq.~\eqref{equ:FPT_majorana_SM}, 
\begin{equation}\label{equ:FPT_Gamma_SM}
    \Gamma^{T_{2}^{f}}=\left(\begin{array}{cc}
        \Gamma^{11} & \text{i}\Gamma^{12}\\
        \text{i}\Gamma^{21} & -\Gamma^{22}
        \end{array}\right),
\end{equation}
where $\Gamma^{bb^\prime}$ represents the block consisting of matrix elements with rows belonging to subsystem $A_b$ and columns belonging to subsystem $A_{b^\prime}$. At the moment, our understanding is limited to the fact that $\Gamma^{T^f_2}$ naturally represents a new Gaussian state, which can generally be expanded using Supplementary Eq.~\eqref{equ:rho_Majorana}:
\begin{equation}
    \rho^\prime=\sum_{\kappa,\tau}w^\prime_{\kappa,\tau}\gamma_{m_{1}}^{\kappa_{1}}\cdots\gamma_{m_{2k}}^{\kappa_{2k}}\gamma_{n_{1}}^{\tau_{1}}\cdots\gamma_{n_{2l}}^{\tau_{2l}}.
\end{equation}
We can further use Wick's theorem for Majorana monomials (products of $2l$ Majorana operators with index different from each other)~\cite{NewJ.Phys.2015Eisler},
\begin{equation}
    \text{Tr}\left(\rho\gamma_{n_{1}}\gamma_{n_{2}}\cdots\gamma_{n_{2l}}\right)=\sum_{\pi}\text{sgn}(\pi)\prod_{k=1}^{l}\Gamma_{n_{\pi(2k-1)},n_{\pi(2k)}},
\end{equation}
to identify the new Gaussian state $\rho^\prime$. Here, $\rho$ is the Gaussian state associated with $\Gamma$, and $\pi$ is a permutation representing different pairs of Majorana operators. On the one hand, we can take $\rho^\prime$ as the Majorana monomial
\begin{equation}
    \text{Tr}\left(\rho\rho^{\prime}\right)=\sum_{\kappa,\tau}w_{\kappa,\tau}^{\prime}\text{Tr}\left(\rho\gamma_{m_{1}}^{\kappa_{1}}\cdots\gamma_{m_{2k}}^{\kappa_{2k}}\gamma_{n_{1}}^{\tau_{1}}\cdots\gamma_{n_{2l}}^{\tau_{2l}}\right).
\end{equation}
On the other hand, we can also use $\rho^\prime$ as the Gaussian state to expand $\rho$
\begin{equation}
    \begin{aligned}
        \text{Tr}(\rho^{\prime}\rho)&=\sum_{\kappa,\tau}w_{\kappa,\tau}\text{Tr}\left(\rho^{\prime}\gamma_{m_{1}}^{\kappa_{1}}\cdots\gamma_{m_{2k}}^{\kappa_{2k}}\gamma_{n_{1}}^{\tau_{1}}\cdots\gamma_{n_{2l}}^{\tau_{2l}}\right)=\sum_{\kappa,\tau}w_{\kappa,\tau}\sum_{\pi}\text{sgn}(\pi)\prod_{p=1}^{k+l}(\Gamma^{T_{2}^{f}})_{\pi(2p-1),\pi(2p)}\\&=\sum_{\kappa,\tau}w_{\kappa,\tau}\text{i}^{|\tau|}\sum_{\pi}\text{sgn}(\pi)\prod_{p=1}^{k+l}\Gamma{}_{\pi(2p-1),\pi(2p)}=\sum_{\kappa,\tau}w_{\kappa,\tau}\text{i}^{|\tau|}\text{Tr}\left(\rho\gamma_{m_{1}}^{\kappa_{1}}\cdots\gamma_{m_{2k}}^{\kappa_{2k}}\gamma_{n_{1}}^{\tau_{1}}\cdots\gamma_{n_{2l}}^{\tau_{2l}}\right).
    \end{aligned}
\end{equation}
Upon comparing the two equations above, we observe that the $\rho^\prime$ linked with $\Gamma^{T^f_2}$ is in fact the partially transposed Gaussian state $\rho^{T^f_2}$.

In short, for each Gaussian state $\rho_\mathbf{s}$ associated with a specific configuration of the auxiliary field, its partial transpose can be expressed via the partial transpose of Green's function in Supplementary Eq.~\eqref{equ:FPT_Gamma_SM}, i.e., $\rho_\mathbf{s}^{T_2^f}=\rho_0[\Gamma^{T_2^f}_\mathbf{s}]$ with $\rho_0[\Gamma]\sim e^{\boldsymbol{\gamma}^\dagger W\boldsymbol{\gamma}}$ and $\tanh(-W/2)=\Gamma$. 

\subsection{Fermionic partial transpose of interacting density matrix}

Finally, by partially transposing the auxiliary-field-dependent Gaussian states in Supplementary Eq.~\eqref{equ:S3_rho_decomp} separately, we obtain the following weighted sum formulation of the partially transposed density matrix:
\begin{equation}\label{equ:S3_rhoT2fMajo}
    \rho^{T_{2}^{f}}=\sum_{{\bf s}}p_{{\bf s}}\rho_{{\bf s}}^{T_{2}^{f}},\quad \rho_{\mathbf{s}}^{T_{2}^{f}}=\det\left[I+e^{W_\mathbf{s}^{T_{2}^{f}}}\right]^{-1/2}e^{\frac{1}{4}\boldsymbol\gamma^{T}W_\mathbf{s}^{T_{2}^{f}}\boldsymbol\gamma},
\end{equation}
where $W_\mathbf{s}^{T_{2}^{f}}=\ln[(I+\Gamma_\mathbf{s}^{T_{2}^{f}})^{-1}(I-\Gamma_\mathbf{s}^{T_{2}^{f}})]$. This formula can be re-expressed in the complex fermion basis, which is more convenient for practical calculations
\begin{equation}
    \rho^{T_{2}^{f}}=\sum_{\bf{s}}p_{\bf{s}}\rho_{\bf{s}}^{T_{2}^{f}},\quad \rho_{{\bf {s}}}^{T_{2}^{f}}=\det\left[G_{{\bf {s}}}^{T_{2}^{f}}\right]\exp\left\{{\mathbf{c}^{\dagger}\ln\left[\left(G_{\mathbf{s}}^{T_{2}^{f}}\right)^{-1}-I\right]\mathbf{c}}\right\},
\end{equation}
where
\begin{equation}\label{equ:S3_Green}
    G^{T_{2}^{f}}=\left(\begin{array}{cc}
        G^{11} & \text{i}G^{12}\\
        \text{i}G^{21} & I-G^{22}
        \end{array}\right). 
\end{equation}

\section{Additional results of the R\'{e}nyi negativity}\label{sec:additional_results}

For a more direct comprehension of the distinction between R\'{e}nyi negativity (RN) and R\'{e}nyi entropy, one can see Supplementary Fig.~\ref{fig:nega-entro1}. In this figure, we compare the R\'{e}nyi entropy $S_2$, the R\'{e}nyi negativity $\mathcal{E}_2$, and the R\'{e}nyi negativity ratio (RNR) $R_2$ for a half-filled Hubbard chain. 
The $S_2-L_{A_1}$ curve in Supplementary Fig.~\ref{fig:nega-entro1}(a) resembles the findings of previous studies (e.g., see Supplementary Ref.~\cite{J.Stat.Mech.2014Broecker}). As the temperature rises, the R\'{e}nyi entropy $S_2$ experiences a quantum-classical transition, shifting from an area law (inclusive of a logarithmic correction) at zero-temperature to a volume law $S_2=L_{A_1}\log 4$ in the high-temperature limit. 
At finite temperatures, while the $S_2-L_{A_1}$ curve exhibits asymmetry about $L_{A_1}=L/2$, the R\'{e}nyi negativity (see Supplementary Fig.~\ref{fig:nega-entro1}(b)) maintains an arc-like structure. Moreover, the two endpoints of the $\mathcal{E}_2-L_{A_1}$ curve correspond to the thermal entropy $S_2^{\mathrm{th}}=S_2(L_{A_1}=L)=\mathcal{E}_2(L_{A_1}=0)=\mathcal{E}_2(L_{A_1}=L)=-\ln(\mathrm{Tr}[\rho^2])$. 
Upon subtracting the thermal entropy from $\mathcal{E}_2$, the negativity ratio $R_2$ (see Supplementary Fig.~\ref{fig:nega-entro1}(c)) emerges as a more capable indicator of finite-temperature entanglement, given its monotonic decrease with rising temperature. 
In the high-temperature limit, the $R_2-L_{A_1}$ curve becomes flat for bulk $L_{A_1}$ values. This plateau can be well described by the formula $\mathcal{E}(L T \gg 1)=\frac{1}{2}\left[\ln \left|\frac{\beta}{\pi} \sinh \left(\frac{\pi L_{A_1}}{\beta}\right)\right|-\frac{\pi L_{A_1}}{\beta}\right]+O\left(\mathrm{e}^{-\pi L T}\right)$ obtained by conformal field theory~\cite{J.Stat.Mech.2019Shapourian} and reflects the area law. 
\begin{figure}[h]
    \centering
    \includegraphics[width=0.8\columnwidth]{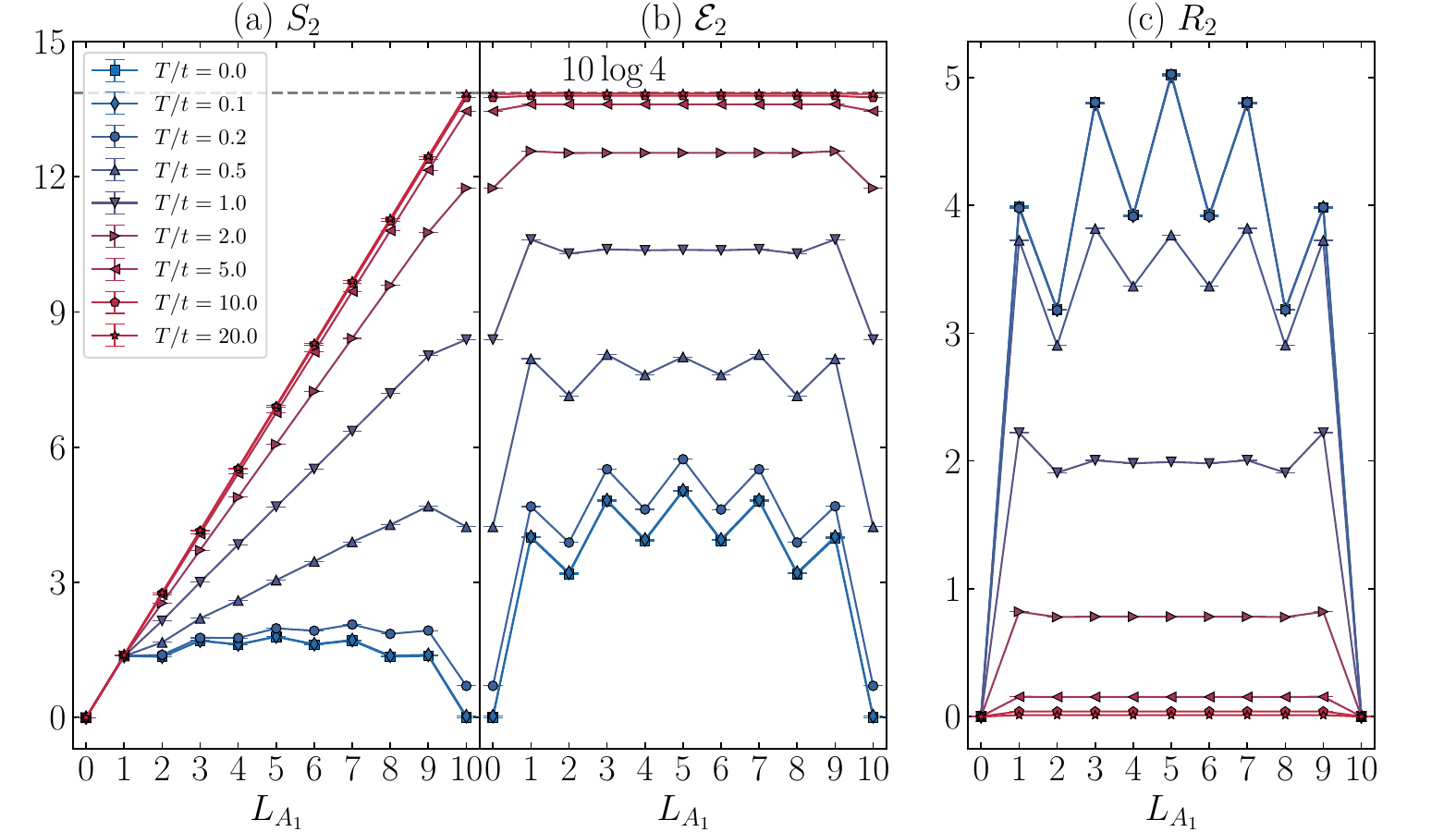}
    \caption{
        \textbf{Comparision of different entanglement measurements for a half-filled Hubbard chain.}
        The comparison of the R\'{e}nyi entanglement entropy $S_2$, the R\'{e}nyi negativity $\mathcal{E}_2$, and the R\'{e}nyi negativity ratio $R_2$ as functions of the subsystem size $L_{A_1}$ is presented for a half-filled Hubbard chain with a length of $L=10(=L_{A_1}+L_{A_2})$ at various temperatures. The dashed line signifies the thermal entropy in the limit as $T$ approaches infinity, represented by $S^{\rm th}_2(T\rightarrow \infty)=L_{A_1}\log 4$~\cite{J.Stat.Mech.2014Broecker}.
        The error bars represent the standard errors from Monte Carlo sampling.
        }
    \label{fig:nega-entro1}
\end{figure}

As shown in Supplementary Fig.~\ref{fig:nega-tvdata}, the super-area-law feature near the finite-temperature critical point of the $t$-$V$ model is evident for two different values of $V$, a different lattice, and a different bipartition geometry. 
Figs.~\ref{fig:nega-tvdata}(a) and (b) show the cases with interaction strengths $V/t=1.0$ and $V/t=0.0$ (free fermion case), respectively. The bipartition geometry remains the same as that in Fig.~3 of the main text and is indicated by the right inset of Supplementary Fig.~\ref{fig:nega-tvdata}(a). In the $V/t=1.0$ case, the peak of $R_2/L$ is located at $T=0.833 t\approx 2T_c$, aligning with prior findings of $T_c/t\approx 0.4\pm 0.1$ obtained by DQMC~\cite{Phys.Rev.B1984Scalapino}. 
In the free fermion case, since $T=0$ is the critical point, the $L\ln L$ scaling is observed at low temperatures, likely associated with the entanglement entropy of the Fermi surface as previous studies on negativity stated~\cite{Phys.Rev.B2016Eisler,J.Stat.Mech.2019Shapourian}. Surprisingly, this scaling persists to a considerable extent at higher temperatures, indicating a prolonged crossover.
Supplementary Fig.~\ref{fig:nega-tvdata}(c) depicts the scenarios with $V/t=2.0$ on a honeycomb lattice. The super-area-law feature is still evident, and the maximum of $R_2/L$ is observed at $T=1.111t\approx 2T_c$, which is slightly higher than the previous result of $T_c/t\approx0.472$ obtained by continuous-time QMC~\cite{Phys.Rev.B2016Hesselmann}.
Supplementary Fig.~\ref{fig:nega-tvdata}(d) presents the case with an equal bipartition geometry on a square lattice. Despite the subpar data quality, the super-area-law feature around the same critical point as the quarter bipartition case is also observed.
\begin{figure}[h]
    \centering
    \includegraphics[width=0.9\columnwidth]{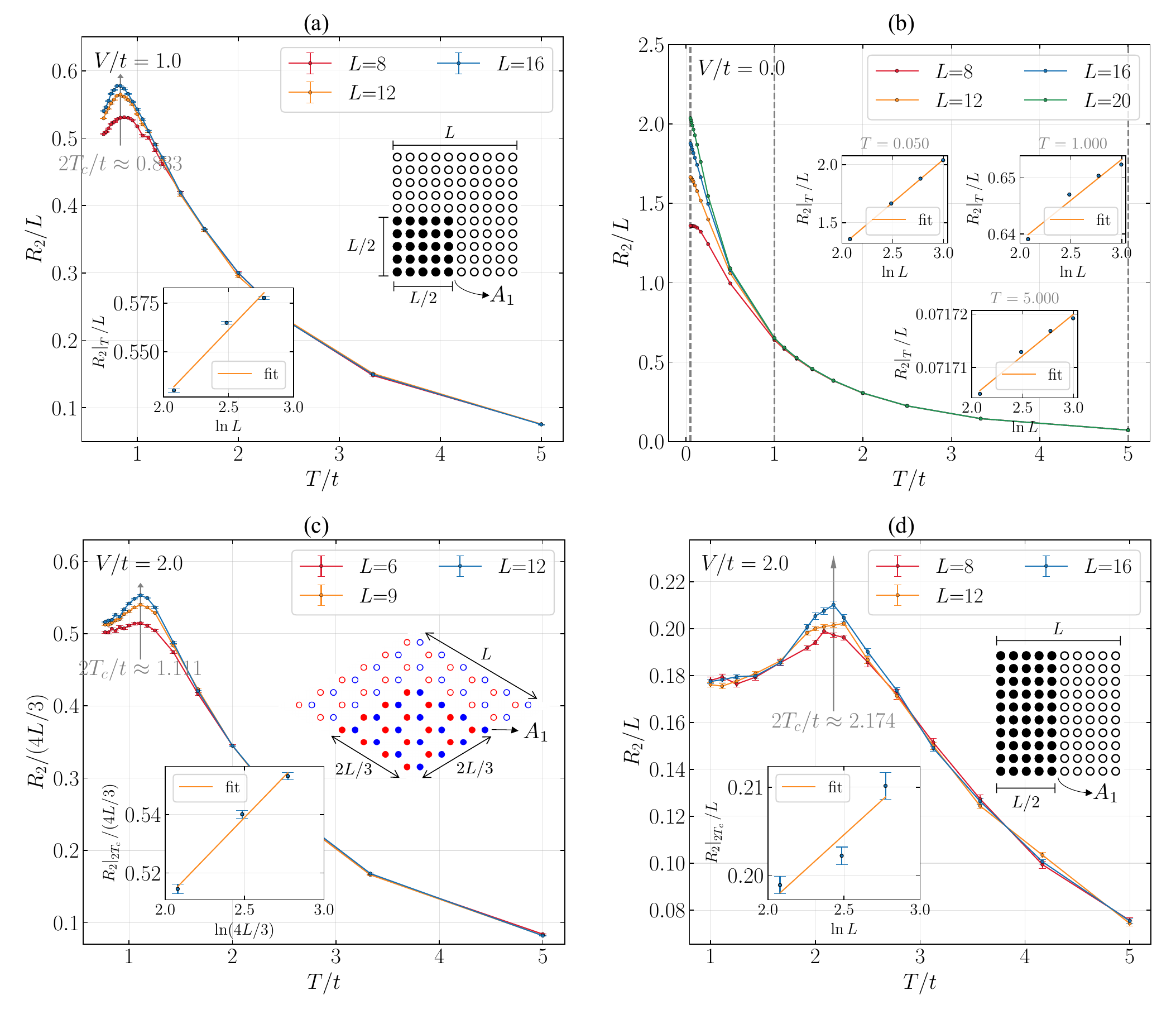}
    \caption{
        \textbf{The beyond-area-law behaviors of the R\'{e}nyi negativity ratio $R_2$ for the $t$-$V$ model in different cases.}
        The area-law coefficients of the R\'{e}nyi negativity ratio $R_2$ as functions of temperature for the $t$-$V$ model. 
        A super-area-law feature around the finite-temperature critical point is evident for (a) $V/t=1.0$, (b) $V/t=0$ (free fermion case), (c) $V/t=2.0$ on a honeycomb lattice and (d) equal-bipartition geometry. In panels (a) and (b), the bipartition geometry matches that of Fig.~3 in the main text, as indicated by the right inset of (a).
        The error bars represent the standard errors from Monte Carlo sampling.
        }
    \label{fig:nega-tvdata}
\end{figure}

\clearpage

\section{Sign problem of Grover determinant}\label{sec:sign_problem}

Using the expression for $\rho_\mathbf{s}^{T^f_2}$, we can calculate rank-$n$ R\'{e}nyi negativity within DQMC framework, 
\begin{equation}
    \begin{aligned}
        e^{-\left(n-1\right)\mathcal{E}_{n}}&= {\rm Tr}\left[\left(\rho^{T_{2}^{f}}\right)^{n}\right] =\sum_{{\bf s}_{1}\dots{\bf s}_{n}}P_{{\bf s}_{1}}\cdots P_{{\bf s}_{n}}\text{Tr}\left[\rho_{{\bf s}_{1}}^{T_{2}^{f}}\cdots\rho_{{\bf s}_{n}}^{T_{2}^{f}}\right]\\&=\sum_{{\bf s}_{1}\dots{\bf s}_{n}}P_{{\bf s}_{1}}\cdots P_{{\bf s}_{n}}\det g_{x}^{n}=\left\langle \det g_{x}^{n}\right\rangle,
    \end{aligned}
\end{equation}
where we have defined the so-called \textit{Grover matrix} $g_x^n$ 
\begin{equation}\label{equ:S4_GroverDeter}
    g_{x}^{n}=G_{{\bf s}_{1}}^{T_{2}^{f}}\cdots G_{{\bf s}_{n}}^{T_{2}^{f}}\left[I+\left(G_{{\bf s}_{1}}^{T_{2}^{f}}\right)^{-1}\left(I-G_{{\bf s}_{1}}^{T_{2}^{f}}\right)\cdots\left(G_{{\bf s}_{n}}^{T_{2}^{f}}\right)^{-1}\left(I-G_{{\bf s}_{n}}^{T_{2}^{f}}\right)\right]
\end{equation}
and its determinant $\det g_x^n$ called \textit{Grover determinant}. As an entanglement measurement, $\mathcal{E}_n\geq 0$ so that $0<\langle \det g_x^n \rangle\leq 1$. It is an interesting question whether for any specific configuration of auxiliary fields $\{\mathbf{s}_1,\dots,\mathbf{s}_n\}$, we always have $\det g_x^n\geq 0$. Moreover, this condition is necessary for the development of an incremental algorithm~\cite{ArXiv2023Liaoa,ArXiv2025Wangb} that can accurately compute R\'{e}nyi negativity, as the weights of all incremental processes include a factor of $(\det g_x^n)^{1/N}$. In this section, we prove that the Grover determinant is real and positive for two classes of sign-problem-free models. We note that these conditions are also applicable to the corresponding Grover determinant associated with entanglement entropy, where all the $G^{T_{2}^{f}}$ in Supplementary Eq.~\eqref{equ:S4_GroverDeter} are replaced by $G^{A_2}$. We only consider models on bipartite lattices and use the notion $(-)^i$ for staggered phase factor that takes $1$ ($-1$) at sites belonging to sublattice $A$ ($B$). 

\subsection{Sufficient Condition I: $G_{ij}^{\downarrow}=(-)^{i+j}(\delta_{ij}-G_{ji}^{\uparrow*})$}

The first class of models includes the half-filled Hubbard model on bipartite lattices. After HS transformation that decouples Hubbard term to density channel as in Supplementary Eq.~\eqref{S2_HSHubbard}, the spacetime-dependent Hamiltonian for a specific configuration of auxiliary fields is given by
\begin{equation}
    H=\sum_i \text{i}C_i(n_{i\uparrow}+n_{i\downarrow}-1)+\sum_{\langle i,j \rangle}D_{ij}(c_{i\uparrow}^{\dagger}c_{j\uparrow}+c_{i\downarrow}^{\dagger}c_{j\downarrow}+\text{h.c.}),
\end{equation}
where $C_i$ and $D_{ij}$ are real constant factors. Turn to a new basis via a partial particle-hole transformation $\tilde{c}_{i\uparrow}= c_{i\uparrow}, \tilde{c}_{i\downarrow}=(-)^i c_{i\downarrow}^{\dagger}$, we obtain
\begin{equation}
    \tilde{H}=\sum_i \text{i}C_i(\tilde{n}_{i\uparrow}-\tilde{n}_{i\downarrow})+\sum_{\langle i,j \rangle}D_{ij}(\tilde{c}_{i\uparrow}^{\dagger}\tilde{c}_{j\uparrow}+\tilde{c}_{i\downarrow}^{\dagger}\tilde{c}_{j\downarrow}+\text{h.c.})
\end{equation}
This Hamiltonian possesses an anti-unitary symmetry $i\sigma_y\mathcal{K}$, where $\sigma_y$ acts on the spin sector and $\mathcal{K}$ means complex conjugate, so it is sign-problem-free. Since the blocks in spin-up and spin-down sectors are complex conjugate to each other, $\tilde{H}_\uparrow=\tilde{H}_\downarrow$, the two blocks in any eigenvector of Hamiltonian and hence the Green's function are also complex conjugate to each other, $\tilde{G}^{\downarrow}_{ij}\equiv\langle\tilde{c}_{i\downarrow}\tilde{c}_{j\downarrow}^{\dagger}\rangle=\tilde{G}^{\uparrow*}_{ij}\equiv\langle\tilde{c}_{i\uparrow}\tilde{c}_{j\uparrow}^{\dagger}\rangle$. Return to the original basis, we find that the Green's functions satisfy the following relation
\begin{equation}\label{equ:S4_suffi1}
    G_{ij}^{\downarrow}=(-)^{i+j}(\delta_{ij}-G_{ji}^{\uparrow*}),
\end{equation}
which can be rewritten as the following matrix form
\begin{equation}
    G^{\downarrow}=U^{\dagger}\left(I-(G^{\uparrow})^{\dagger}\right)U
\end{equation}
with $U_{ij}=\delta_{ij}(-)^i=\delta_{ij}(-)^j$ a diagonal unitary matrix. 

The condition in Supplementary Eq.~\eqref{equ:S4_suffi1} is sufficient for $\det g_x^n\geq 0$. Consider the spin-up block of partially transposed Green's function in Supplementary Eq.~\eqref{equ:S3_Green}, which satisfies
\begin{equation}
    U^{\dagger}\left(G^{\uparrow,T_{2}^{f}}\right)^{\dagger}U=VG^{\downarrow,T_{2}^{f}}V+I,\text{ with } V=\left(\begin{array}{cc}
        \text{i}I_{1}\\
         & -\text{i}I_{2}
        \end{array}\right). 
\end{equation}
$V$ is a diagonal unitary matrix satisfying $V^2=-I$. Reformulate the above relation we have
\begin{equation}
    \begin{aligned}
        G^{\downarrow,T_{2}^{f}}&=V^{\dagger}U^{\dagger}\left(I-\left(G^{\uparrow,T_{2}^{f}}\right)^{\dagger}\right)UV,\\I-G^{\uparrow,T_{2}^{f}}&=UV\left(G^{\downarrow,T_{2}^{f}}\right)^{\dagger}V^{\dagger}U^{\dagger},\\1-G^{\downarrow,T_{2}^{f}}&=V^{\dagger}U^{\dagger}\left(G^{\uparrow,T_{2}^{f}}\right)^{\dagger}UV.
    \end{aligned}
\end{equation}
Using the above relations, one can prove that $\det g_x^n=\det g_x^{n,\uparrow}\det g_x^{n,\downarrow}\geq 0$ since
\begin{equation}
    \begin{aligned}
        \det g_{x}^{n,\downarrow}\equiv&\det\left\{ G_{{\bf s}_{1}}^{\downarrow,T_{2}^{f}}\cdots G_{{\bf s}_{n}}^{\downarrow,T_{2}^{f}}\left[I+\left(G_{{\bf s}_{1}}^{\downarrow,T_{2}^{f}}\right)^{-1}\left(I-G_{{\bf s}_{1}}^{\downarrow,T_{2}^{f}}\right)\cdots\left(G_{{\bf s}_{n}}^{\downarrow,T_{2}^{f}}\right)^{-1}\left(I-G_{{\bf s}_{n}}^{\downarrow,T_{2}^{f}}\right)\right]\right\} \\=&\det\left\{ \left[I+G_{{\bf s}_{n}}^{\uparrow,T_{2}^{f}}\left(I-G_{{\bf s}_{n}}^{\uparrow,T_{2}^{f}}\right)^{-1}\cdots G_{{\bf s}_{1}}^{\uparrow,T_{2}^{f}}\left(I-G_{{\bf s}_{1}}^{\uparrow,T_{2}^{f}}\right)^{-1}\right]\left(I-G_{{\bf s}_{n}}^{\uparrow,T_{2}^{f}}\right)\cdots\left(I-G_{{\bf s}_{1}}^{\uparrow,T_{2}^{f}}\right)\right\} ^{*}\\=&\det\left\{ G_{{\bf s}_{1}}^{\uparrow,T_{2}^{f}}\cdots G_{{\bf s}_{n}}^{\uparrow,T_{2}^{f}}\left[I+\left(G_{{\bf s}_{1}}^{\uparrow,T_{2}^{f}}\right)^{-1}\left(I-G_{{\bf s}_{1}}^{\uparrow,T_{2}^{f}}\right)\cdots\left(G_{{\bf s}_{n}}^{\uparrow,T_{2}^{f}}\right)^{-1}\left(I-G_{{\bf s}_{n}}^{\uparrow,T_{2}^{f}}\right)\right]\right\} ^{*}\\\equiv&\left(\det g_{x}^{n,\uparrow}\right)^{*}.
        \end{aligned}
\end{equation}

\subsection{Sufficient Condition II: $\Gamma_{ij}^{(2)}=(-)^{i+j}\Gamma_{ij}^{(1)*}$}

The second class of models, including the spinless $t$-$V$ model on bipartite lattices, are proved in Majorana basis. For convenience, let us first relabel the Majorana operators by introducing a specie index, i.e., we use $\gamma^{(1)}_i$ and $\gamma^{(2)}_i$ to represent $\gamma_{2i-1}$ and $\gamma_{2i}$, respectively. After HS transformation that decouples NN interaction terms to Majorana hopping channel as in Supplementary Eq.~\eqref{S2_HStV}, the spacetime-dependent Hamiltonian for a specific configuration of auxiliary fields is given by
\begin{equation}
    H=\sum_{\left\langle ij\right\rangle }C_{ij}\left({\rm i}\gamma_{i}^{\left(1\right)}\gamma_{j}^{\left(1\right)}+{\rm i}\gamma_{i}^{\left(2\right)}\gamma_{j}^{\left(2\right)}\right),
\end{equation}
where $C_i$ are real constant factors. This Hamiltonian possesses an anti-unitary symmetry $T\mathcal{K}$, where $T$ transforms $\gamma_i^{(1)}$ ($\gamma_i^{(2)}$) to $(-)^i \gamma_i^{(2)}$ ($(-)^i \gamma_i^{(1)}$), so it is sign-problem-free. Turn to a new basis via transformation $\tilde{\gamma}_{i}^{(1)}=\gamma_{i}^{(1)}, \tilde{\gamma}_{i}^{(2)}=(-)^i\gamma_{i}^{(2)}$, the Hamiltonian becomes
\begin{equation}
    \tilde{H}=\sum_{\left\langle ij\right\rangle }C_{ij}\left({\rm i}\tilde{\gamma}_{i}^{\left(1\right)}\tilde{\gamma}_{j}^{\left(1\right)}-{\rm i}\tilde{\gamma}_{i}^{\left(2\right)}\tilde{\gamma}_{j}^{\left(2\right)}\right).
\end{equation}
Since the coefficients of $ \tilde{\gamma}_{i}^{\left(1\right)}\tilde{\gamma}_{j}^{\left(1\right)} $ and $ \tilde{\gamma}_{i}^{\left(2\right)}\tilde{\gamma}_{j}^{\left(2\right)} $ are complex conjugate to each other, using similar arguments as the case of Hubbard model, we have the relation $\tilde{\Gamma}^{(2)}=\tilde{\Gamma}^{(1)*}$. Return to the original basis, we find that Green's functions satisfy the following relation
\begin{equation}\label{equ:S4_suffi2}
    \Gamma_{ij}^{(2)}=(-)^{i+j}\Gamma_{ij}^{(1)*}\text{ or }\Gamma^{(2)}=U^\dagger\Gamma^{(1)*}U.
\end{equation}

The condition stated in Supplementary Eq.~\eqref{equ:S4_suffi2} is also sufficient for $\det g_x^n\geq 0$. According to the expression of partially transposed Green's function in Supplementary Eq.~\eqref{equ:FPT_Gamma_SM} and $W^{T^f_2}=\ln[(I+\Gamma^{T^f_2})^{-1}(I-\Gamma^{T^f_2})]$, we can derive the relationships between the two blocks pertaining to two distinct Majorana species sectors:
\begin{equation}
    \Gamma^{(2),T_{2}^{f}}=J^{\dagger}U^{\dagger}\left(\Gamma^{\left(1\right)T_{2}^{f}}\right)^{*}UJ \text{ and }W^{(2),T_{2}^{f}}=J^{\dagger}U^{\dagger}\left(W^{\left(1\right),T_{2}^{f}}\right)^{*}UJ\text{ with }J=\left(\begin{array}{cc}
        I_{1}\\
         & -I_{2}
        \end{array}\right).
\end{equation}
Write the Grover determinant in Majorana basis using Supplementary Eq.~\eqref{equ:S3_rhoT2fMajo}, and it can be easily proved that $\det g_x^n=\det g_x^{n,(1)}\det g_x^{n,(2)}\geq 0$ since
\begin{equation}
    \begin{aligned}
        \det g_{x}^{n,(2)}=&\det\left[I+e^{W_{{\bf s}_{1}}^{\left(2\right),T_{2}^{f}}}\right]^{-1/2}\cdots\det\left[I+e^{W_{{\bf s}_{n}}^{\left(2\right),T_{2}^{f}}}\right]^{-1/2}\det\left[I+e^{W_{{\bf s}_{1}}^{\left(2\right),T_{2}^{f}}}\cdots e^{W_{{\bf s}_{n}}^{\left(2\right),T_{2}^{f}}}\right]^{1/2}\\=&\det\left[I+e^{W_{{\bf s}_{1}}^{\left(1\right),T_{2}^{f},*}}\right]^{-1/2}\cdots\det\left[I+e^{W_{{\bf s}_{n}}^{\left(1\right),T_{2}^{f},*}}\right]^{-1/2}\det\left[I+e^{W_{{\bf s}_{1}}^{\left(1\right),T_{2}^{f},*}}\cdots e^{W_{{\bf s}_{n}}^{\left(1\right),T_{2}^{f},*}}\right]^{1/2}\\=&\det g_{x}^{n,(1)*}.
        \end{aligned}
\end{equation}
\bibliography{reference_nega.bib}